\documentclass[twocolumn,prd,a4paper,final,superscriptaddress,longbibliography,nofootinbib]{revtex4-1}

\usepackage{mathrsfs} 
\usepackage{amsmath,amssymb,amsfonts}
\usepackage{color}
\usepackage{graphicx}
\usepackage{bbold}
\usepackage[T1]{fontenc}
\usepackage{epsfig}
\usepackage{epstopdf}
\usepackage{bbold}
\usepackage{comment}
\usepackage{bm}
\newcommand{\gnabla}{\ooalign{\hfil/\hfil\crcr$\nabla$}}

\usepackage{natbib}

\graphicspath{{./figax/}}

\begin{document}
	
	\title{Soliton solution of a gravitating chiral quark soliton model: dynamical fermion and Dirac-sea in general relativity}
	
	\author{Ryoma Mizutani}
	\affiliation{Department of Physics and Astronomy, Tokyo University of Science, Noda, Chiba 278-8510, Japan}
	
	\author{Daiki Niiyama}
	\affiliation{Department of Physics and Astronomy, Tokyo University of Science, Noda, Chiba 278-8510, Japan}
	
	\author{Nobuyuki Sawado}
	\email{sawadoph@rs.tus.ac.jp}
	\affiliation{Department of Physics and Astronomy, Tokyo University of Science, Noda, Chiba 278-8510, Japan}

	\vspace{.5 in}
	\small

	\date{\today}
	
	\begin{abstract}
	In this paper, we study Einstein-Dirac system based on a Dirac fermion 
	coupled with a nonlinear chiral field on static spherically symmetric spacetime. 
	Gravitational effects on the dynamical mass of fermions are examined.
	The chiral quark soliton model (CQSM) originally was a model of hadron inspired by the low-energy 
	regime of large-$N_\textrm{c}$ QCD,  
	realizing localized fermions with the full inclusion of the Dirac-sea, which is derived 
	from a regularized one-fermion loop. 
	We extend the CQSM in spherically symmetric curved spacetime and coupled with the Einstein gravity.  
	We successfully solve the CQSM and the Einstein equation self-consistently, 
	and obtain the spectral flow of the fermion energies and also of the ADM mass. 
	The Dirac-sea dilutes the effect of the energy-momentum tensor, which induces an inactive point for 
	the impact of gravity even when the localizing matter exists. The Dirac-sea effect becomes dominant 
	for larger dynamical mass $M\sim \langle\bar{\psi}\psi\rangle$, leading to the emergence of 
	the negative ADM mass.
	\end{abstract}
	

	\maketitle 
	\section{\label{Intro}Introduction}
	
	The Einstein-Dirac (ED) system is a system of fermions minimally coupled with Einstein gravity~\cite{Deser:1974cy}.
Finster \textit{et al}.~\cite{Finster:1998ws} demonstrated the existence of solutions with systems in spherically 
symmetric space-time, and numerous kinds of solutions have been investigated 
up to this point~\cite{Herdeiro:2017fhv,Bronnikov:2019nqa, Dzhunushaliev:2018jhj,daRocha:2020rda,
Leith:2020jqw,Leith:2021urf, BakuczCanario:2020qmq,Iwasawa:2025owk,Zhang:2025jco}.
The Pauli exclusion principle is inherently responsible for keeping Dirac spinor structures from collapsing under 
their own gravity. Especially, in \cite{Leith:2020jqw,Leith:2021urf, Iwasawa:2025owk}, the authors 
discussed existence of the many fermion states in spherically symmetric curved space-time. 

Numerous studies have investigated the Dirac fermions in curved space-time,
coupling of Dirac fermions with electromagnetism~\cite{Finster:1998jqw}, and  
Yang-Mills fields~\cite{Finster:2000vy} (see also recent follow-up studies  
in~\cite{Dzhunushaliev:2019kiy,Dzhunushaliev:2019uft,Blazquez-Salcedo:2020czn,Kain:2023ann,
Kain:2023pvp,Dzhunushaliev:2023ylf}.) 
Among these, \cite{Dzhunushaliev:2024kti} gives us a new insight for mechanism 
of the localization of the Dirac fermions, 
where the localization of the fermions in Schwarzschild-like space-time is 
caused by so called the hedgehog type solution with the topological charge $Q=1$ gravitating-skyrmions. 
The topology is an essential role 
for the property and the typical spectral flow of the fermions in such coupled system 
has been extensively studied. 

A Dirac Hamiltonian should have eigenstates of the Dirac-sea, a collection of particles with negative energy, 
as well as single particle (valence) states. At a Dirac equation coupled with topological solitons, 
the normalizable zero modes exist corresponding to the valence fermions. The concept of the Dirac-sea 
is an old story but is still useful for understanding stability of the vacuum with Dirac fermions.
A vacuum without fermions in negative energy state is unstable in the sense that, every fermion can decay
into the negative energy state. Therefore, the empty negative energy states should be filled for preventing 
such decay.
On phenomenology, 
the effect of the Dirac-sea has been studied in the context of quark nuclear matter
in the strong magnetic field; the enhancement of the light quark condensate with increase in external 
magnetic field is observed~\cite{Mizher:2010zb}  
and also for heavier quark sector~\cite{Mishra:2022hqb}.    
For the nuclear matter, 
effect of the Dirac-sea on hadronic properties in SU(3) chiral invariant model~\cite{Mishra:2003tr} and
the phase transition of hot and dense matter~\cite{Mukherjee:2018ebw} 
have been discussed in several types of models.   
Also, the spin symmetry in antiproton and antineutron or in the Dirac-sea 
have been discussed~\cite{Zhou:2003iu,Shen:2018irh}. 
Formulation of the one-loop effective action naturally implements zero-point energy fluctuation 
or the Dirac-sea~\cite{Dashen:1975xh} as a Casimir energy. The vacuum contributions of the fermions coupled with
several solitonic objects such as kink~\cite{Weigel:2023fxe, Saadatmand:2022htx}, 
the cosmic string~\cite{Quandt:2017hbu,Weigel:2015lva,Graham:2011fw}, 
and, for example, the non-abelian strings~\cite{Weigel:2009wi}.
The method for the one-loop effective action also is very useful for the problem of field theoretical 
model in curved space-times~\cite{Parker:2009uva,Parker:1983pe,Flachi:2014jra,Flachi:2010yz,Flachi:2011zr}. 
The Nambu-Jona-Lasinio model in curved space-time has been the subject of many papers 
~\cite{Hill:1991jc,Elizalde:1994zv,Inagaki:1993ya,Inagaki:1997kz,Addazi:2017qus}.
The model we will employ in the present paper can be regarded as a semi-bosonized version of 
this famous model. 

We study localizing, normalizable solutions of fermion including Dirac-sea in a spherically symmetric, 
Schwarzschild-like space-time based on the chiral quark soliton model (CQSM). 
The CQSM was developed in 
1980's as a low-energy effective theory of Quantum Chromodynamics (QCD). 
Since the model includes the Dirac-sea quark 
contribution and explicit valence quark degrees of 
freedom, it successfully interpolates between the 
constituent quark model and the Skyrme 
model~\cite{Diakonov:1987ty,Reinhardt:1989st,Meissner:1990tz,Wakamatsu:1990ud,Christov:1995vm,Alkofer:1995mv,Diakonov:2000pa}. 
The Dirac-sea is realized in terms of appropriately regularized one-loop 
effective action, and the stable soliton of only the quark degrees of freedom is 
generated by a balance between the Dirac-sea and the valence quarks.
The CQSM incorporates the non-perturbative
feature of the low-energy QCD, spontaneous breaking of the chiral symmetry and generation of the dynamical mass of fermion. 
It has been shown that, after a proper quantization scheme such like a collective coordinate quantization method is applied,  
the $B=1$ solution provides correct observable 
such as a nucleon including mass, electromagnetic value, spin carried
by quarks, parton distributions and octet
SU(3) baryon spectra~\cite{Blotz:1992pw,Kim:1995mr,Kim:1995ha,Christov:1995hr,Wakamatsu:1997en,Wakamatsu:1998rx,Dressler:1999zg,Goeke:2000wv}. 
Subsequently, the model demonstrated a crucial role in so-called \textit{the penta-quark}. 
To the best of our knowledge, only the so-called valence (positive energy)  fermions have been shown to exist as localizing fermions in terms of gravity and there is no concrete analysis that implements the Dirac-sea.
This is our central concern in the present paper. 

The remainder of this paper is organized as follows. 
Section \ref{sec2} briefly introduce the Einstein-Dirac system and derives the Schwarzschild-like metric, 
the gamma matrices and the coupled Dirac and Einstein equations. 
The numerical method is over-viewed in Sec.~\ref{sec3}. 
Section \ref{sec4} provides successful solutions and energies of the system.
Concluding remarks are presented in Sec.~\ref{sec5}.

	\section{\label{sec2}The chiral quark soliton model as a Einstein-Dirac system}
	
	\subsection{The Einstein-Dirac system}
	We first give a basic review the Einstein-Dirac system, which
	has been extensively studied in recent years~\cite{Herdeiro:2017fhv,Bronnikov:2019nqa, Dzhunushaliev:2018jhj,daRocha:2020rda,
Leith:2020jqw,Leith:2021urf, BakuczCanario:2020qmq,Iwasawa:2025owk,Zhang:2025jco,
Dzhunushaliev:2019kiy,Dzhunushaliev:2019uft,Blazquez-Salcedo:2020czn,Kain:2023ann,Kain:2023pvp,Dzhunushaliev:2023ylf,
Dzhunushaliev:2024kti}. 
Throughout this paper, 
	we adopt a mostly minus $(+,-,-,-)$ metric signature. The action of the 
	Einstein-Dirac system comprises the Einstein-Hilbert action $S_\textrm{G}$ and also a
	standard Dirac action $S_\textrm{D}$ which is given by
	\begin{align}
	&S=
	-\int d^4x\sqrt{-g}\frac{\mathcal{R}}{16\pi G}+S_\textrm{D}
	\label{action}
	\end{align}
	where $\mathcal{R}$ is the Ricci scalar and the $G$ is the gravitational constant and 
	$g\equiv\textrm{det}(g_{\mu\nu})$ 
	where $g_{\mu\nu}(x)$ is the metric of the space-time. 
	A most commonly used $S_\textrm{D}$ for the Dirac spinor $\Psi$ is of course  
	a free Lagrangian $\mathcal{L}_\textrm{D}=\bar{\Psi}(i\gnabla-m)\Psi$, where 
	the fermion current mass denoted by $m$ is expected to be small one. 
	The covariant derivative $\gnabla=\underline{\gamma}^\mu(\partial_\mu-\Gamma_\mu)$ is written in 
	terms of the well-known spin connection $\Gamma_\mu$ implementing effects of the gravity 
	into the fermion~\cite{Dolan:2015eua}
	and also the gamma matrices $\underline{\gamma}^\mu$ in curved space-time, satisfying Clifford algebra 
	\begin{align}	
	\{\underline{\gamma}^\mu,\underline{\gamma}^\nu\}=2g^{\mu\nu}\,.
	\end{align}	
	Extremizing the action \eqref{action}
	with respect to the metric and Dirac field yields the Einstein and the Dirac equations. 
	The coupled equations can easily be solved
	in some numerical algorithm, e.g.,the shooting method. 
	 
	We assume the resulting object will be of spherically symmetric and the metric is defined by 
	spherical coordinates $(t,r,\theta,\phi)$ as
	\begin{align}
	ds^2=\sigma^2(r)N(r)dt^2-\frac{dr^2}{N(r)}-r^2(d\theta^2+\sin^2\theta d\phi^2).
	\label{lineelement}
	\end{align}
	The fields $\sigma (r), N(r)$ asymptotically connect to the well-known Schwarzschild metric as follows:
	\begin{align}
	\sigma(r)\to 1,~~N(r)\to 1-\frac{2GM_\textrm{ADM}}{r}
	\end{align}
	where $M_\textrm{ADM}$ is the Arnowitt-Deser-Misner mass. 
	Inserting the ansatz into the Einstein-Hilbert action yields the simple reduced action~
	\cite{Breitenlohner:1991aa,Breitenlohner:1994di,Dzhunushaliev:2024kti}
	\begin{align}
	S^\textrm{red}_\textrm{G}=-\frac{1}{2G}\int dtdr\sigma(r)(N-1+r N')\,.
	\label{redEH}
	\end{align}
	
	For constructing Dirac Lagrangian in curved space-time, we start with the vierbein and the 
	spin connections. The vierbein $e^\mu_a$ 
	is defined in terms of
	\begin{align}
	g_{\mu\nu}=e_{a\mu}e^a_\nu,~~\eta_{ab}=e_{a\mu}e^\mu_b
	\end{align}
	where $\eta_{ab}=\textrm{diag}(1,-1,-1,-1)$. The gamma-matrices in curved space-time $\underline{\gamma}^\mu$
	are defined in terms of the normal (flat) $\gamma$-matrices $\gamma^\mu$ 
	as follows:
	\begin{align}
	\underline{\gamma}^\mu=e^\mu_a\gamma^a
	\end{align}
	Of course the flat $\gamma^\mu$ should satisfy the Clifford algebra $\{\gamma^\mu,\gamma^\nu\}=\eta^{\mu\nu}$.
	In terms of \eqref{lineelement}, the vierbein is explicitly written as
	\begin{align}
&e_a^\mu=
\begin{pmatrix}
\frac{1}{\sigma\sqrt{N}}&0&0&0\\
0&\sqrt{N}\sin\theta\cos\phi&\sqrt{N}\sin\theta\sin\phi&\sqrt{N}\cos\theta\\
0&\frac{1}{r}\cos\theta\cos\phi&\frac{1}{r}\cos\theta\sin\phi&-\frac{1}{r}\sin\theta\\
0&-\frac{1}{r\sin\theta}\sin\phi&\frac{1}{r\sin\theta}\cos\phi&0\\
\end{pmatrix} 
\nonumber \\
&\hspace{3cm}a=(0,1,2,3),\mu=(t,r,\theta,\phi)\,,
\end{align}
and the curved $\gamma$-matrices in polar coordinate as 
\begin{align}
\underline{\gamma}^t&=\frac{1}{\sigma\sqrt{N}}\gamma^0\,,
\label{gamma0}\\
\underline{\gamma}^r&=\sqrt{N}\left(\gamma^1\sin\theta\cos\phi+\gamma^2\sin\theta\sin\phi+\gamma^3\cos\theta\right)\,,
\\
\underline{\gamma}^\theta&=\frac1r\left(\gamma^1\cos\theta\cos\phi+\gamma^2\cos\theta\sin\phi-\gamma^3\sin\theta\right)\,,\\
\underline{\gamma}^\phi&=\frac1{r\sin\theta}\left(-\gamma^1\sin\phi+\gamma^2\cos\phi\right)\,.
\label{gamma3}
\end{align}
For definition of the spin connections 
\begin{align}
\Gamma_\mu:=-\frac{1}{8}{\omega_\mu^a}_b[\gamma_a,\gamma^b]
=\frac{1}{8}e_b^\nu (\partial_\mu e^a_\nu-\Gamma^\lambda_{\mu\nu}e^a_\lambda)[\gamma_a,\gamma^b]\,,
\end{align}
the explicit form are summarized as
\begin{align}
&\Gamma_t=-\frac{1}{4}(\sigma^2N)'\underline{\gamma}^t\underline{\gamma}^r\,,
\label{spinconnection0}\\
&\Gamma_r=0\,,
\\
&\Gamma_\theta=-\frac{r}{2}\biggl(\frac{1}{\sqrt{N}}-1\biggr)\underline{\gamma}^\theta\underline{\gamma}^r\,,
\\
&\Gamma_\phi=-\frac{r\sin^2\theta}{2}\biggl(\frac{1}{\sqrt{N}}-1\biggr)\underline{\gamma}^\phi\underline{\gamma}^r\,.
\label{spinconnection3}
\end{align}
In the present paper, we employ the following Dirac representation of the flat $\gamma$-matrices
\begin{align}
\gamma^0&=
\begin{pmatrix}
1&0\\
0&-1\\
\end{pmatrix},~~
\gamma^i&=
\begin{pmatrix}
0 &\sigma^i \\
-\sigma^i & 0 \\
\end{pmatrix} ,~~~~i=1,2,3.
\label{flatgamma}
\end{align}
	
	Finster \textit{et al.}~\cite{Finster:1998ws,Finster:1998jqw} numerically solved 
	the problem of two gravitationally localized neutral fermions, 
assuming opposite spins to ensure spherical symmetry. 
Leith \textit{et al.}~\cite{Leith:2020jqw,Leith:2021urf} studied the  
many fermion systems, arranging $N_\textrm{f}$th fermions 
in a filled shell of total angular momentum $j=\frac{N_\textrm{f}-1}{2}$. 
We have extended their formulation into the multi-shell model which certainly stabilizes the system~\cite{Iwasawa:2025owk}. 

In addition to studies of noninteracting free Dirac fermions,  
several interacting models have been studied. The simple four-point interactions enjoying the nonlinearity 
were considered for increasing the total mass 
~\cite{Bronnikov:2019nqa,Dzhunushaliev:2018jhj,daRocha:2020rda,Zhang:2025jco} 
\begin{align}
\mathcal{L}_\textrm{int}=-\frac{\lambda}{2}(\bar{\Psi}\Psi)^2\,.
\end{align}
Particularly, localized fermions coupled with gravitating-skyrmions~\cite{Dzhunushaliev:2024kti}
is closely related to our study in the present paper. 
The interaction Lagrangian is defined by the following chiral coupling scheme
\begin{align}
\mathcal{L}_\textrm{int}=-M\bar{\Psi}U^{\gamma_5}\Psi,~~
U^{\gamma_5}:=\frac{1+\gamma_5}{2}U+\frac{1-\gamma_5}{2}U^\dagger
\end{align}
where the coupling constant $M$ is so called the constituent mass or the dynamical mass 
which realizes the mass-gap of the fermions, and 
$U:=\exp (i\bm{\tau}\cdot\bm{\pi}/f_\pi)$ is the chiral field in terms of the isotriplet 
pion fields $\bm{\pi}=(\pi_1,\pi_2,\pi_3)$, and also $\gamma_5:=i\gamma^0\gamma^1\gamma^2\gamma^3$. 
The topological solitons $U$ in 3+1 dimensions 
are the well-known skyrmion configuration. 
The Lagrangian of the chiral quark soliton model (CQSM) in the curved space-time
is therefore defined by the following chiral invariant form 
\begin{align}
\mathcal{L}_\textrm{D}=\bar{\Psi}(i\gnabla-MU^{\gamma_5})\Psi
\end{align}
where $\gnabla:=\underline{\gamma}^\mu(\partial_\mu-\Gamma_\mu)$. 
The axial coupling constant (the pion decay constant) $f_\pi =93$ MeV is essential quantity 
of physics of the spontaneous symmetry breaking of chiral invariant models. 

For convenience, we introduce dimensionless quantities. 
The length scale of the model parameters and fields are
\begin{align}
\Psi:[L^{-3/2}],~~U:[L^0],~~M:[L^{-1}]\,.
\label{field}
\end{align}
Since $f_\pi$ has a dimension of inverse of the length 
scale in the natural unit,   
\eqref{field} are transformed into the dimensionless quantities $\tilde{x},\tilde{\Psi},\tilde{M}$ 
\begin{align}
&x^\mu\to \tilde{x}^\mu:=(a f_\pi) x^\mu,~~\Psi\to \tilde{\Psi}:={(af_\pi)}^{-3/2}\Psi,
\nonumber \\
&M\to \tilde{M}:={(af_\pi)}^{-1}M
\end{align}
where $a$ is some dimensionless constant. 
In the following, we omit $"\tilde{~~}"$ for simplicity. 
Additionally, for use in the Einstein equations later on, 
we define a dimensionless effective gravitational coupling 
constant $\alpha^2:=2 G(af_\pi)^2N_c$
\footnote{In this paper, we set $af_\pi=200.0$ MeV, but the analysis 
does not require its use unless you want to see the solutions' true radius or  dimensionful energy.}.

\subsection{\label{ssec23}Gravitating chiral quark soliton model}

The chiral quark soliton model (CQSM) is a well-known low-energy effective model of Quantum Chromodynamics
~\cite{Diakonov:1987ty,Reinhardt:1989st,Meissner:1990tz,Wakamatsu:1990ud,Christov:1995vm,Alkofer:1995mv,Diakonov:2000pa}. 
The model is formally defined by an effective partition function, and
in this paper we extend the model into the curve space-time as
\begin{align}
\mathcal{Z}&=\int\mathscr{D}\Psi\mathscr{D}\Psi^\dagger \mathscr{D}U
\exp\biggl[i\int\sqrt{-g}d^4x\bar{\Psi}(i\overleftrightarrow{\gnabla}-MU^{\gamma_5})\Psi\biggr]
\nonumber \\
&=\int \mathscr{D}U \exp\Bigl(iS_\textrm{eff}[U]\Bigr)
\label{effectiveaction_flat}
\end{align}
where we have employed an abbreviated notation
$\overleftrightarrow{\gnabla}:=\underline{\gamma}^\mu(\overleftrightarrow{\partial}_\mu-\Gamma_\mu)$, which
guarantees the Hermiticity of the corresponding Hamiltonian.  
The symbol $\overleftrightarrow{\partial}$ implies 
$\bar{\psi}\overleftrightarrow{\partial}\phi=\frac{1}{2}(\bar{\psi}\partial \phi-\partial\bar{\psi}\phi)$.  
After integrating the fermion fields, we obtain the effective action
\begin{align}
S_\textrm{eff}[U]&=
-iN_\textrm{c}\log \det iD
\nonumber \\
&=-iN_\textrm{c}\textrm{Sp}\log iD,~~iD:=i\overleftrightarrow{\gnabla}-MU^{\gamma_5}
\label{action1}
\end{align}
where the trace Sp contains the trace of the spin tr$_\gamma$ and the flavor tr$_f$ such as
\begin{align}
\textrm{Sp}\hat{O}=\int dt\textrm{Tr}\hat{O}:=\int\sqrt{-g}dtd^3x\textrm{tr}_\gamma\textrm{tr}_f\langle x|\hat{O}|x\rangle\,.
\nonumber 
\end{align}
In order to reduce complexity, we eliminate $\overleftrightarrow{~~}$ from the formalism in the following, 
yet the differential operator always involves the function.
In this paper, we employ the spectral method for explicitly deriving the effective action, i.e., 
Eq.\eqref{action1} is expressed via the eigenvalues of the Dirac equation coupled with the skyrmions and 
also the metrics. Since all of the background fields, including the metrics and skyrmions, are static, 
the approach is the most suitable for solving the current problem.

In flat space-time, the Dirac operator $iD_\textrm{f}$ can be written as
\begin{align}
iD_\textrm{f}=\gamma^0(i\partial_t-H)
\label{Diracopflat}
\end{align}
where the Hamiltonian $H$ is defined as
\begin{align}
H:=-i\gamma^0\gamma^i\partial_i+\gamma^0MU^{\gamma_5}\,.
\end{align}
The determinant in Eq.\eqref{action1} is thus derived in terms of 
the eigenvalues of the Schr\"odinger operator~\cite{Dashen:1975xh,Reinhardt:1989st}
\begin{align}
(i\partial_t-H)\Psi_{\nu,n}=\lambda_{\mu,n}\Psi_{\mu,n}
\end{align}
where $\Psi_{\nu,n}$ have to satisfy anti-periodic boundary condition with the period $T$
\begin{align}
\Psi_{\mu,n}(\bm{x},t+T)=-\Psi_{\mu,n}(\bm{x},t)\,.
\end{align}
The eigenvalues are 
\begin{align}
\lambda_{\mu,n}=-\epsilon_\mu+\frac{2n+1}{T}\pi,~~~~n=0,1,2,\cdots\,.
\end{align} 
The determinant is evaluated 
\begin{align}
&\det(\gamma^0(i\partial_t-H))=\det (i\partial_t-h)=\prod_{\mu,n}\lambda_{\mu,n}
\nonumber \\
&=C\exp[i\frac{T}{2}\sum_\mu|\epsilon_\mu|]
\prod_\mu(1+\exp(-iT|\epsilon_\mu|))
\end{align}
where $C=\prod_{\mu,n>0}(-1)((2n+1)\pi/T)^2$. 
Therefore the effective action reads
\begin{align}
S_\textrm{eff}=N_\textrm{c}T\sum_{\mu}\biggl(\frac{|\epsilon_\mu|}{2}-n_\mu|\epsilon_\mu|\biggr)
\label{action_flat}
\end{align}
where $n_\mu$ means occupation numbers of one-particle (valence) fermion. The action implements the 
vacuum Dirac-sea as well as the one-particle and anti-particle contribution for the definite set of 
occupation numbers. 
Eq.\eqref{action_flat} is formally divergent thus some regularization must be required for physically meaningful
answer. 

In the curved space-time~\eqref{lineelement}, we express the Dirac operator 
in terms of the gamma matrices \eqref{gamma0}-\eqref{gamma3} 
and the spin-connection \eqref{spinconnection0}-\eqref{spinconnection3} as  
\begin{align}
&i\tilde{D}=i\underline{\gamma}^t\partial_t+i\underline{\gamma}^r \mathcal{D}_r
+i\underline{\gamma}^\theta\partial_\theta+i\underline{\gamma}^\phi\partial_\phi
-MU^{\underline{\gamma}_5}\,,
\\
&\mathcal{D}_r:=\partial_r+\frac{1}{r}\biggl(1-\frac{1}{\sqrt{N}}\biggr)
+\frac{1}{4}(\log\sigma^2N)'\,.
\end{align}
It can be verified that the matrix $\underline{\gamma}_5$ in the curved 
space-time coincides with the flat space-time~\cite{Dzhunushaliev:2024kti}.
To develop the analysis, 
it is convenient to rewrite the Dirac operator into a form similar to Eq.\eqref{Diracopflat} 
in a flat space-time~\cite{Parker:1980hlc,Parker:1980kw,Leclerc:2005wj,Antoine:2019fwz} 
\begin{align}
i\tilde{D}:=\underline{\gamma}^t(i\partial_t-\tilde{H})
\end{align}
where $\tilde{H}$ is the Dirac Hamiltonian in curved space defined by
\begin{align}
\tilde{H}&=(g^{tt})^{-1}\underline{\gamma^t}[-i(\underline{\gamma}^r\mathcal{D}_r
+\underline{\gamma}^\theta\partial_\theta+\underline{\gamma}^\phi\partial_\phi)
+MU^{\underline{\gamma}_5}]\,.
\label{Hamiltonian}
\end{align} 
Also, we introduce a operator $\tilde{D}_0$ which is $\tilde{D}$ with $U=1$ (vacuum)
\begin{align}
&i\tilde{D}_0=\underline{\gamma}^t(i\partial_t-\tilde{H}_0)\,,
\nonumber \\
&\tilde{H}_0=(g^{tt})^{-1}\underline{\gamma^t}[-i(\underline{\gamma}^r\mathcal{D}_r
+\underline{\gamma}^\theta\partial_\theta+\underline{\gamma}^\phi\partial_\phi)
+M]\,.
\label{DiracHamiltonian0}
\end{align}
We define the scalar product for the wave functions $\Psi,\Phi$ defined to be~\cite{Parker:1980hlc}
\begin{align}
(\Psi,\Phi)=\int d^3x\sqrt{-g}\Psi^\dagger(x)\gamma^0\underline{\gamma}^t\Phi(x)\,.
\end{align}
For the static background, the Dirac spinor 
$\Psi(x)$ can be separated into the time and the space components
\begin{align}
\Psi(\bm{x},t)=\phi_\mu(\bm{x})e^{-i\varepsilon_\mu t}
\end{align}
and then, the Dirac eigenequation is
\begin{align}
\tilde{H}\phi_\mu(\bm{x}) = \varepsilon_{\mu}\phi_\mu(\bm{x})\,.
\label{DiracEigen}
\end{align}
These eigenstates and also the equation of the vacuum Hamiltonian \eqref{DiracHamiltonian0}
is
\begin{align}
\tilde{H}_0\phi^0_k(\bm{x})=\varepsilon_k\phi^0_k(\bm{x})
\label{DiracEigen0}
\end{align}
are used to estimate the effective action~\cite{Reinhardt:1989st}. 
In the metric \eqref{lineelement}, 
the orthonormal condition for the Dirac spinor $\phi_\mu(\bm{x})$ becomes 
\begin{align}
(\phi_\mu,\phi_\nu)=\int d^3x\frac{1}{\sqrt{N}}\phi_\mu^\dagger(\bm{x})\phi_\nu(\bm{x})=\delta_{\mu\nu}\,.
\label{orthogonality}
\end{align}
As in Eq.\eqref{DiracEigen}, we numerically solve the 
vacuum Dirac equation \eqref{DiracEigen0} because it contains the 
metric function $\sigma,N$ (otherwise, the eigenvalues is just $\varepsilon=\pm\sqrt{k^2+M^2}$). 
We will discuss in detail the numerical method in the later section. 
The effective action \eqref{action1} is a complex valued quantity 
and we concentrate on the following real part
\begin{align}
S_\textrm{eff}&
=-i\frac{N_\textrm{c}}{2}\textrm{Sp}[\log \tilde{D}^\dagger \tilde{D}-\log \tilde{D}_0^\dagger \tilde{D}_0]
\nonumber \\
&=-i\frac{N_\textrm{c}}{2}\textrm{Sp}[\log D^\dagger D-\log D_0^\dagger D_0]\,.
\label{efaction}
\end{align}
Here we introduce slightly modified Dirac operators
\begin{align}
&iD=\gamma^0(i\partial_t-\tilde{H}),~~
iD_0=\gamma^0(i\partial_t-\tilde{H}_0)
\end{align}
for convenience. 
Eq.\eqref{efaction} is still a divergent object and then needs to be regularized. The 
proper-time regularization~\cite{Schwinger:1951nm} is used
\begin{align}
S^\textrm{reg}_\textrm{eff}=i\frac{N_\textrm{c}}{2}\int_{1/\Lambda^2}^\infty\frac{d\tau}{\tau} \textrm{Sp}
[e^{-D^\dagger D\tau}-e^{-D_0^\dagger D_0\tau}]\,.
\end{align}
Since the theory is not normalizable, we oblige to remain the cutoff parameter $\Lambda$, which is 
determined such that the derivative expansion of the above effective action reproduce the correct 
coefficient of the kinetic term in the nonlinear sigma model~\cite{Wakamatsu:1990ud}.
This requirement leads to the condition
\begin{align}
f_\pi^2=\frac{N_\textrm{c}M^2}{4\pi^2}\int_{1/\Lambda^2}^\infty \frac{d\tau}{\tau}e^{-\tau M^2}\,.
\end{align} 
When we choose $f_\pi=93$ MeV, for the typical value of the dynamical mass $M=400.0$ MeV 
the value of the cutoff parameter is $\Lambda=636.0$ MeV, 
or, for the dimensionless mass $M=2.0$, the cutoff is $\Lambda=3.18$. 
However, using a different option of $f_\pi$ may still result in a different value of $\Lambda$.

For explicit calculation of $\partial_t^2$ in Eq.\eqref{efaction}, 
we introduce a normalized eigenfunction 
satisfying the anti-periodic boundary condition with the period $[0,T]$, i.e.
$\langle\omega_n|t\rangle=T^{-1/2}\exp(-i\omega_nt)$. 
The temporal trace can be evaluated such that
\begin{align}
&\textrm{Sp}f(i\partial_t)=\int dt\langle t|\textrm{Tr}f(i\partial_t)|t\rangle
\nonumber \\
&=\sum_n\langle\omega_n|\textrm{Tr}f(\omega_n)|\omega_n\rangle
\to T\int\frac{dz}{2\pi}\textrm{Tr}f(z)\,.
\end{align}
Note that Tr contains the trace over the space, Dirac, and flavor indices. 
For the explicit evaluation of the fermion determinant, it is reasonable to switch to
Euclidean space by Wick rotation such as $z\to iz$. The result is 
\begin{align}
S^\textrm{reg}_\textrm{eff}=-T\frac{1}{2}\frac{N_\textrm{c}}{\sqrt{4\pi}}
\int_{1/\Lambda^2}^\infty \frac{d\tau}{\tau^{3/2}}\textrm{Tr}[e^{-\tau \tilde{H}^2}-e^{-\tau \tilde{H}_0^2}]\,.
\label{effaction}
\end{align}
The Tr in Eq.\eqref{effaction} can be derived in terms of the eigenstates $\phi_\mu(\bm{x})$ of 
Eq.\eqref{DiracEigen} and also the those of the unperturbed $U=1$ Hamiltonian \eqref{DiracEigen0} $\phi_k^0(\bm{x})$. 
The result defines the reduced one-loop effective action 
\begin{align}
S^\textrm{red}_\textrm{sea}:=
-\frac{1}{2}\frac{N_\textrm{c}}{\sqrt{4\pi}}
\int_{1/\Lambda^2}^\infty \frac{d\tau}{\tau^{3/2}}\biggl(\sum_\mu e^{-\tau \varepsilon_\mu^2}-\sum_k e^{-\tau \varepsilon_k^2}\biggr)
\label{ractionvac}
\end{align}
which describes the Dirac-sea contribution to the total action.  
Thanks to well-known the Index Theorem, there is a normalizable zero-mode when the Dirac spinor interacts to
the skyrmion with the charge $B=1$. 
We define action of the valence (one particle) fermion
\begin{align}
S^\textrm{red}_\textrm{val}:=-N_\textrm{c}\sum_\mu n_\mu\theta(\varepsilon_\mu)\varepsilon_\mu
\label{ractionval}
\end{align}
where $\theta(\varepsilon_\mu)$ is a Heaviside step function and 
$n_\mu=0,1$ is the valence fermion occupation number. 
In Sec.\ref{sec4}, we shall present result of the spectral flow in the present system which
connects the valence fermion number and the topological charge $B$ of the skyrmion.  
For $B=1$, only the ground state with the eigenvalue 
$\varepsilon_\textrm{val}\equiv \varepsilon_0$, 
which comes from upper continuum is occupied, i.e.,
$n_0=1$ and all the others are zero. 
When the ground state descends into the negative energy $\varepsilon_\textrm{val}< 0$, 
the vacuum contribution \eqref{ractionvac} carries the one-fermion contribution 
since the sums over $\mu$ in Eq.\eqref{ractionvac} span all single-particle levels, 
both positive and negative~\cite{Meissner:1991tv}. 
The similar mechanism repeatedly implements validity of 
the formalism enjoying both the valence and the vacuum Dirac-sea contributions, such as
the total energy~\eqref{fermionenergy}, the density of the fermion~\eqref{fermiondensity}, 
the skyrmion equation~\eqref{eqm_skyrme}, 
and also the source terms of the Einstein equations~\eqref{EMtensors2},\eqref{EMtensorN2}.  
Accordingly, we define the reduced Dirac effective action
\begin{align}
S_\textrm{D}^\textrm{red}:=S^\textrm{red}_\textrm{val}+S^\textrm{red}_\textrm{sea}\,.
\label{ractionD}
\end{align}

\begin{figure*}[t]
	\centering 
	\includegraphics[width=0.5\linewidth]{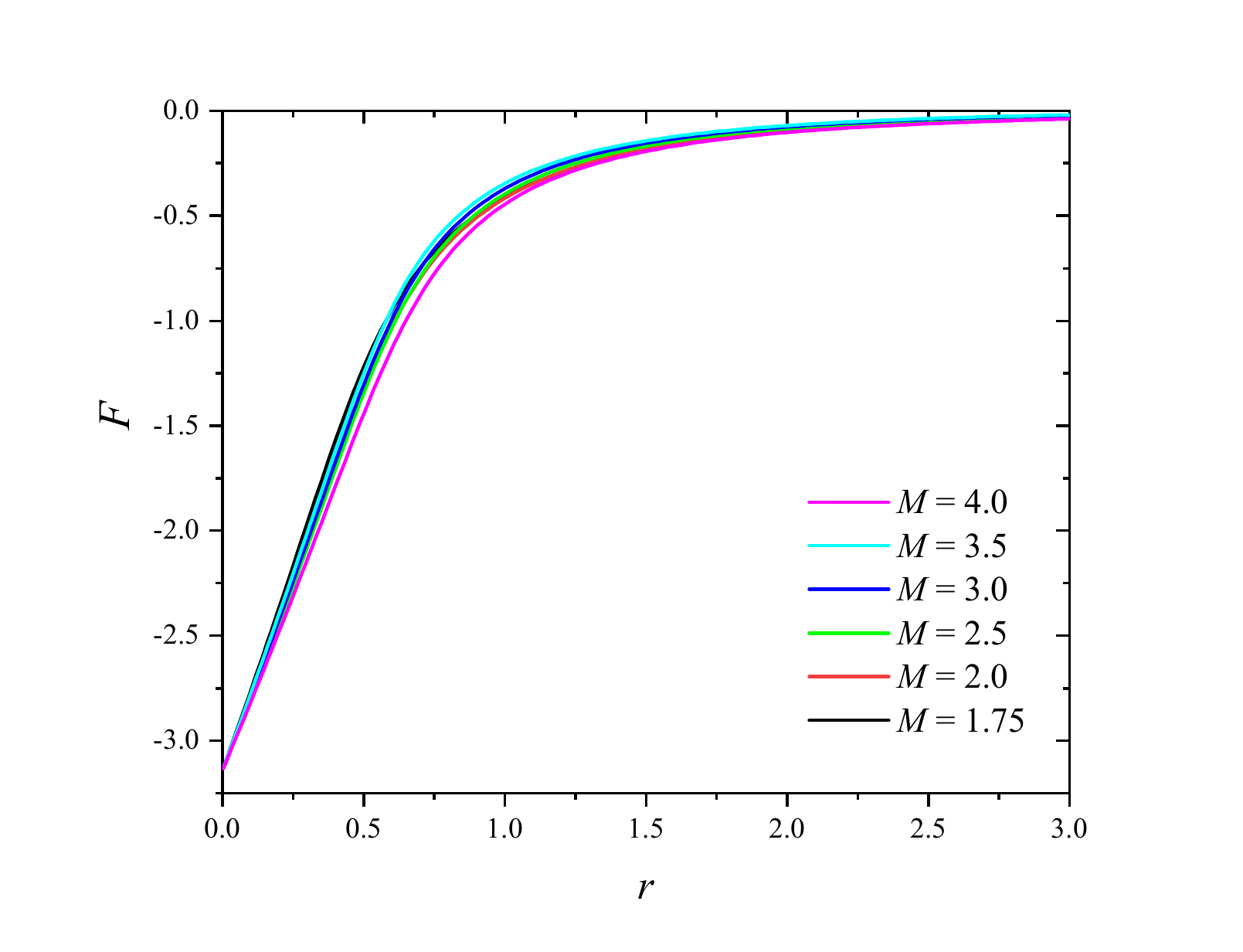}\hspace{-1.0cm}
	\includegraphics[width=0.5\linewidth]{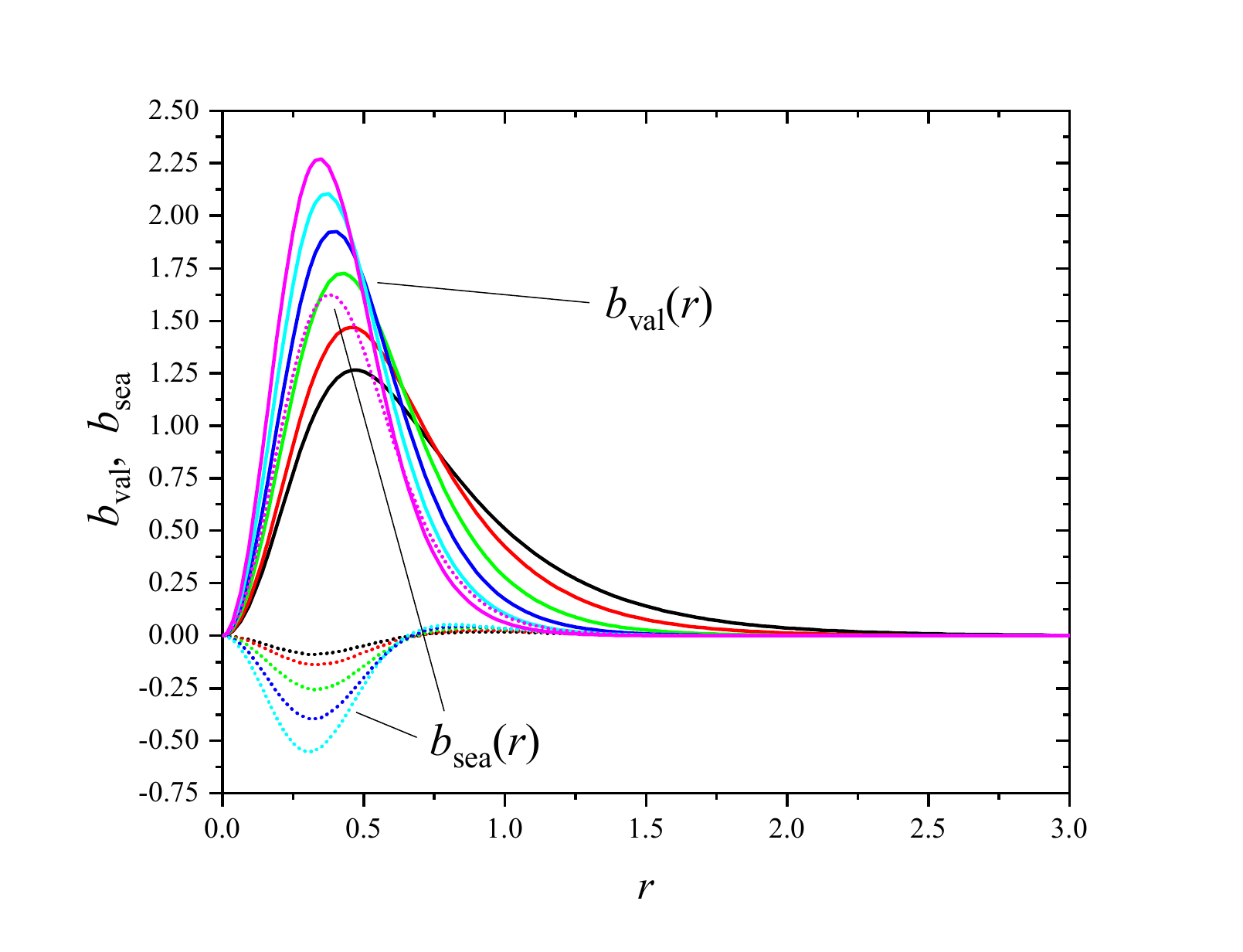}

	\caption{\label{Solution_flat}The self-consistent flat ($\alpha^2=0$) solutions 
	for several dynamical mass $M=1.75, 2.0, 2.5, 3.0, 3.5$ and $4.0$. 
	The right figure is the profile function $F(r)$ 
	and the left is the valence and the sea fermion density $b_\textrm{val}(r),b_\textrm{sea}(r)$, 
	defined by \eqref{fermiondensity}. 
	Both figures share the same color chart of the lines.
	Since the spectrum of the solution with $M=4.0$ dives into the negative energy region, 
	the Dirac-sea contribution has the positive value. }
\end{figure*}


Since the gravitational field is described by a static, spherically symmetric metric \eqref{lineelement}, 
it is quite natural to implement the spherically symmetric ansatz both for the skyrmion $U$ and also the Dirac spinor. 
In this paper, we employ a solution of the hedgehog form for the skyrmion profile
\begin{align}
U(\bm{x})=\exp(i\bm{\tau}\cdot\hat{\bm{r}}F(r))
\end{align}
where the profile function $F(r)$ is depends only on the radial coordinate $r$.  
By imposing the boundary condition for $F$ such that
\begin{align}
F(0)=-\pi,~~F(\infty)=0\,,
\label{spinorBC}
\end{align}
the resulting skyrmion has the topological charge $Q=1$. 
For the action $S_\textrm{D}^\textrm{red}$, 
we impose the extremum condition
\begin{align}
\frac{\delta}{\delta F(r)}S^\textrm{red}_\textrm{D}=0
\end{align}
which leads to the equation of motion for the Skyrme profile:
\begin{align}
&S(r)\sin F(r)=P(r)\cos F(r)\,,
\\
&S(r):=\sum_\mu (n_\mu\theta(\varepsilon_\mu)+\mathcal{N}(\varepsilon_\mu)\textrm{sgn}(\varepsilon_\mu))
\nonumber \\
&\hspace{1cm}\times\bigl(\phi_\mu(\bm{x}),(g^{tt})^{-1}\underline{\gamma^t}
\delta(|\bm{x}|-r)\phi_\mu(\bm{x})\bigr)\,,
\nonumber \\
&P(r):=\sum_\mu (n_\mu\theta(\varepsilon_\mu)+\mathcal{N}(\varepsilon_\mu)\textrm{sgn}(\varepsilon_\mu))
\nonumber \\
&\hspace{1cm}
\times \bigl(\phi_\mu(\bm{x}),i(g^{tt})^{-1}\underline{\gamma^t}~\underline{\gamma}_5
\bm{\tau}\cdot\hat{\bm{r}}\delta(|\bm{x}|-r)\phi_\mu(\bm{x})\bigr)
\nonumber 
\label{eqm_skyrme}
\end{align}
where
\begin{align}
\mathcal{N}(\varepsilon_\mu):=-\frac{1}{\sqrt{4\pi}}\Gamma\biggl(\frac{1}{2}, \Bigl(\frac{\varepsilon_\mu}{\Lambda}\Bigr)^2\biggr)\,.
\end{align}

We construct the reduced action consists of the reduced Einstein-Hilbert action \eqref{redEH} 
and the CQSM action \eqref{ractionD} so that
\begin{align}
S^\textrm{red}:=S^\textrm{red}_\textrm{G}+S^\textrm{red}_\textrm{D}\,.
\end{align}
Varying $S^\textrm{red}$ with respect to $\sigma,N$, 
we obtain the coupled Einstein equations:
\begin{align}
N-1+rN'&=-\alpha^2T_\sigma[\sigma,N]\,,
\label{Einsteins} \\
r\sigma'&=\alpha^2T_N[\sigma,N]\,.
\label{EinsteinN}
\end{align}
The dimensionless gravitational coupling constant 
$\alpha^2 =2 G(af_\pi)^2N_\textrm{c}$ was already introduced in the previous section. 
The definition may be instructive because it implies the system
is composed of gravity ($"G"$), weak  ($"f_\pi"$) and strong interaction ($"N_c"$).
The source terms in the right hand side $T_\sigma,T_N$ are defined in terms of 
the Dirac eigenstates as follows
\begin{align}
&T_\sigma[\sigma,N]:=T_\sigma[\sigma,N]_\textrm{val}+T_\sigma[\sigma,N]_\textrm{sea}\,,
\nonumber \\
&T_\sigma[\sigma,N]_\textrm{val}=\sum_\mu n_\mu  \theta(\varepsilon_\mu)
\frac{\delta}{\delta\sigma(r)}\bigl(\phi_\mu(\bm{x}),\tilde{H}\phi_\mu(\bm{x})\bigr)\,,
\nonumber \\
&T_\sigma[\sigma,N]_\textrm{sea}=\sum_\mu\textrm{sgn}(\varepsilon_\mu)\mathcal{N}(\varepsilon_\mu)
\frac{\delta}{\delta\sigma(r)}\bigl(\phi_\mu(\bm{x}),\tilde{H}\phi_\mu(\bm{x})\bigr)
\nonumber \\
&\hspace{1.0cm}-\sum_k\textrm{sgn}(\varepsilon^0_k)\mathcal{N}(\varepsilon^0_k)\frac{\delta}{\delta\sigma(r)}
\bigl(\phi^0_k(\bm{x}),\tilde{H}_0\phi^0_k(\bm{x})\bigr)\,;
\label{EMtensors2}
\\
&T_N[\sigma,N]:=T_N[\sigma,N]_\textrm{val}+T_N[\sigma,N]_\textrm{sea}\,,
\nonumber \\
&T_N[\sigma,N]_\textrm{val}=\sum_\mu n_\mu  \theta(\varepsilon_\mu)
\frac{\delta}{\delta N(r)}\bigl(\phi_\mu(\bm{x}),\tilde{H}\phi_\mu(\bm{x})\bigr)\,,
\nonumber \\
&T_N[\sigma,N]_\textrm{sea}=\sum_\mu\textrm{sgn}(\varepsilon_\mu)\mathcal{N}(\varepsilon_\mu)
\frac{\delta}{\delta N(r)}\bigl(\phi_\mu(\bm{x}),\tilde{H}\phi_\mu(\bm{x})\bigr)
\nonumber \\
&\hspace{1.0cm}-\sum_k\textrm{sgn}(\varepsilon^0_k)\mathcal{N}(\varepsilon^0_k)\frac{\delta}{\delta N(r)}
\bigl(\phi^0_k(\bm{x}),\tilde{H}_0\phi^0_k(\bm{x})\bigr)\,.
\label{EMtensorN2}
\end{align}

\begin{figure}[t]
	\centering 
	\includegraphics[width=1.0\linewidth]{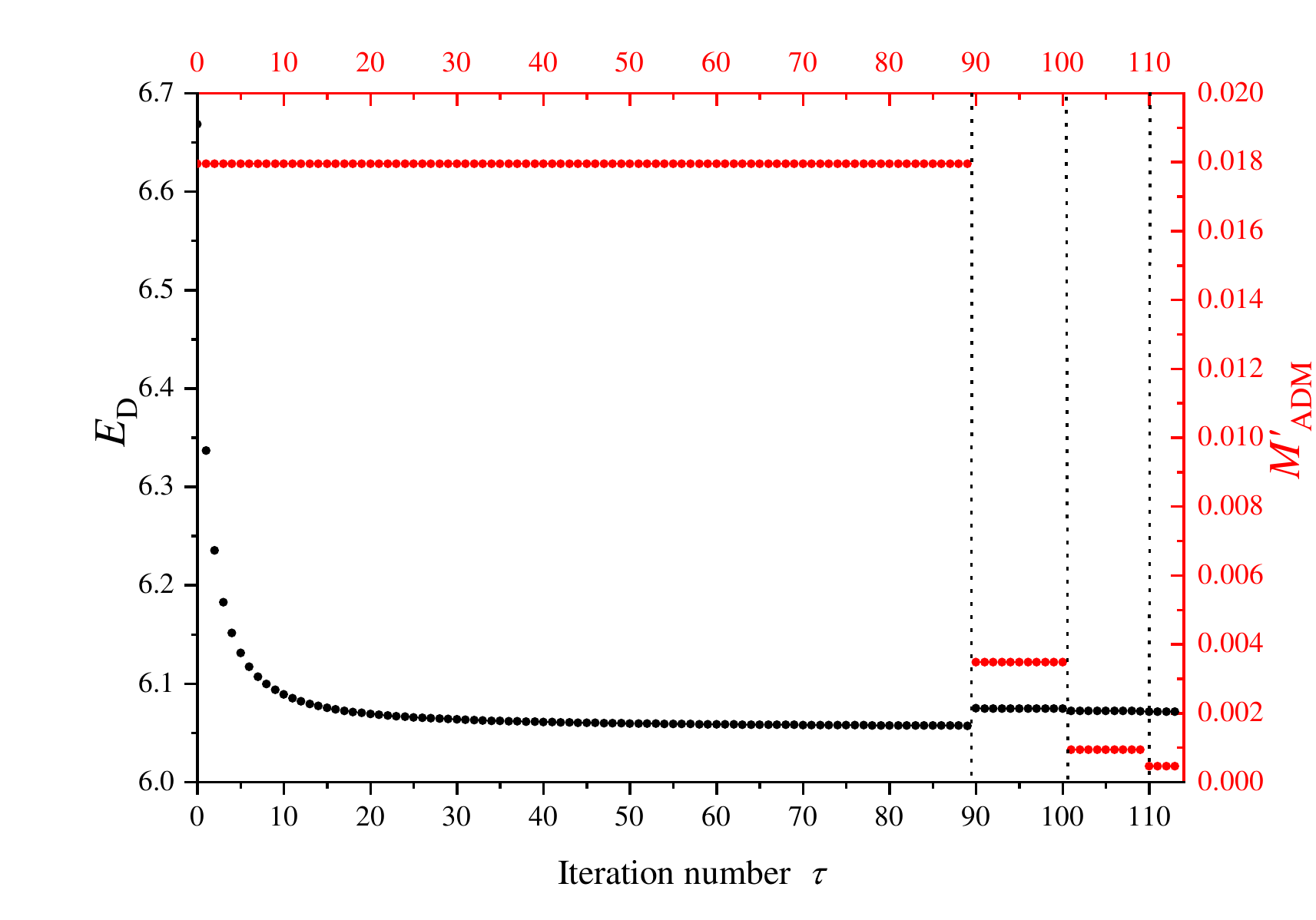}

	\caption{\label{Iteration}The numerical iteration process of the total fermion energy $E_D$  
	and 	the ADM mass $M'_\textrm{ADM}$ as a iteration step $\tau$ for
	the dynamical mass $M=2.0$ and the coupling constant $\alpha=0.1$.
	The first sequence is $0\leq \tau \leq 89$, 
	where the simulation is performed for $\sigma_0(r) =N(r)=1.0$ until the convergence is achieved. 
	Then, the second sequence $90\leq \tau \leq 100$ is performed 
	for the metric functions at $\tau=89$ evaluated via Eqs.\eqref{metricfunctions}. 
	The third sequence is $101  \leq \tau \leq 109$ and the fourth is $110 \leq \tau \leq 113$. 
	After these steps, the Dirac eigenstates, 
	the profile function, and the metric functions are evaluated at each iteration step.}
\end{figure}

\section{\label{sec3}Numerical Method}

In order to find the solutions of the present system, we have to solve different types of equations such 
as the Dirac equations \eqref{DiracEigen} and \eqref{DiracEigen0}, 
the equation of motion of the Skyrme profile~\eqref{eqm_skyrme}, 
and the Einstein equations \eqref{Einsteins} and \eqref{EinsteinN}.
In general, the coupled equations like the ED system can be solved directly to get solutions. 
However, when the Dirac-sea is considered, the problem becomes more complex and cannot be solved simultaneously.
Instead, we explore the self-consistent solutions of the present system. 
\\
\textbf{Step 1:}~ Assuming an initial profile $F_0(r)$ satisfying the boundary condition \eqref{spinorBC}
and metrics $\sigma_0(r),N_0(r)$, we solve the Dirac equations 
\eqref{DiracEigen} and \eqref{DiracEigen0}. 
\\
\textbf{Step 2:}~ The functions $S(r),P(r)$ in Eq.\eqref{eqm_skyrme} are investigated in terms of the obtained eigenvalues 
and the eigenstates. The algebraic equation of motion~\eqref{eqm_skyrme} makes it simple to obtain the new profile function $F(r)$. 
\\
\textbf{Step 3:}~Using the eigenvalues and eigenstates, the energy-momentum tensors $T_\sigma(r),T_N(r)$ are examined 
in Eqs.\eqref{Einsteins} and \eqref{EinsteinN}. 
After fixing the energy-momentum tensor, we simply integrate the Einstein equations~\eqref{Einsteins} and \eqref{EinsteinN} 
to determine the new metrics $\sigma (r), N(r)$. 
Plugging these solutions $F(r),\sigma(r),N(r)$ into $F_0(r),\sigma_0(r),N_0(r)$, we repeat the \textbf{Step 1}.  
\\
The procedure \textbf{Step 1} - \textbf {Step 3} is repeated until the self-consistency of the solutions is attained. 

In the following subsections, we describe the explicit method of each equation.

\begin{figure}[t]
	\centering 
	\includegraphics[width=1.0\linewidth]{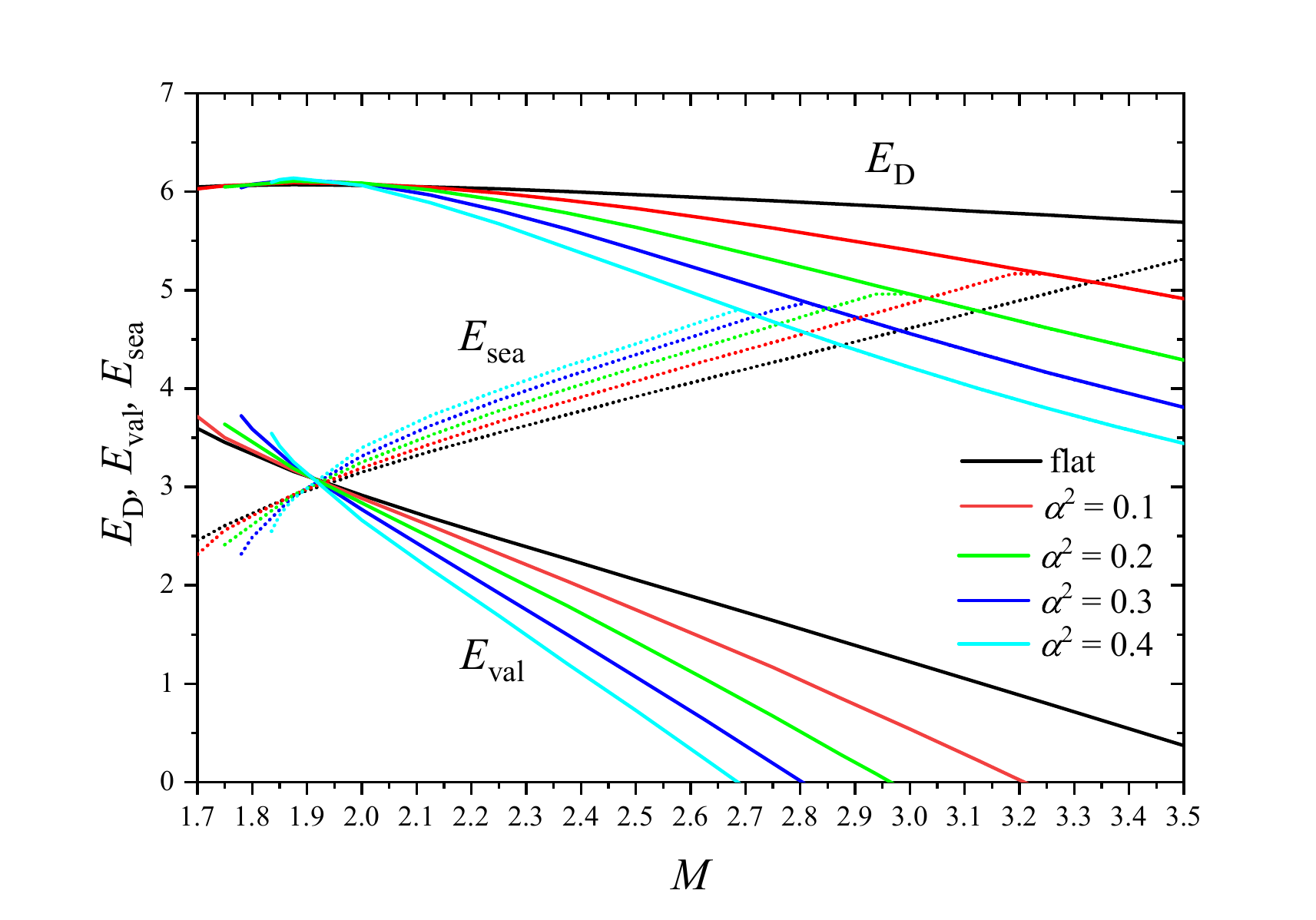}

	\caption{\label{Spectral flow}The valence, the Dirac-sea and their sum of the energy of the 
	fermions \eqref{fermionenergy} with the dynamical mass $M:[1.7,3.5]$ for the several gravitational 
	coupling constant $\alpha^2=0,0.1,0.2,0.3,0.4$ is plotted. 
	When $M$ increases, the valence energy falls into the negative region, and the Dirac-sea component 
	carries the total energy. The number of color is $N_c=3$.	
	}
\end{figure}

\begin{figure*}[htbp]
	\centering 
	\includegraphics[width=0.5\linewidth]{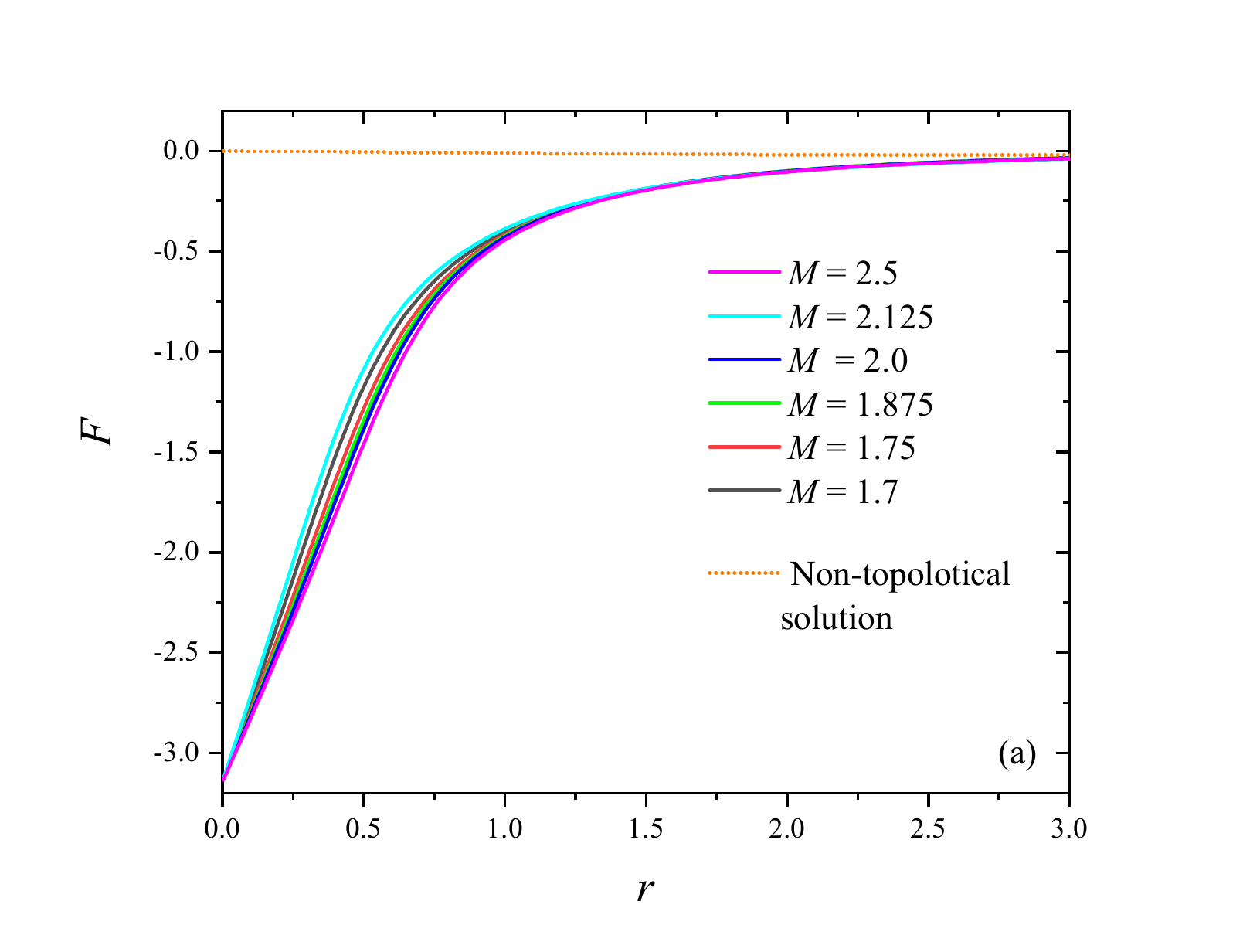}\hspace{-1.0cm}
	\includegraphics[width=0.5\linewidth]{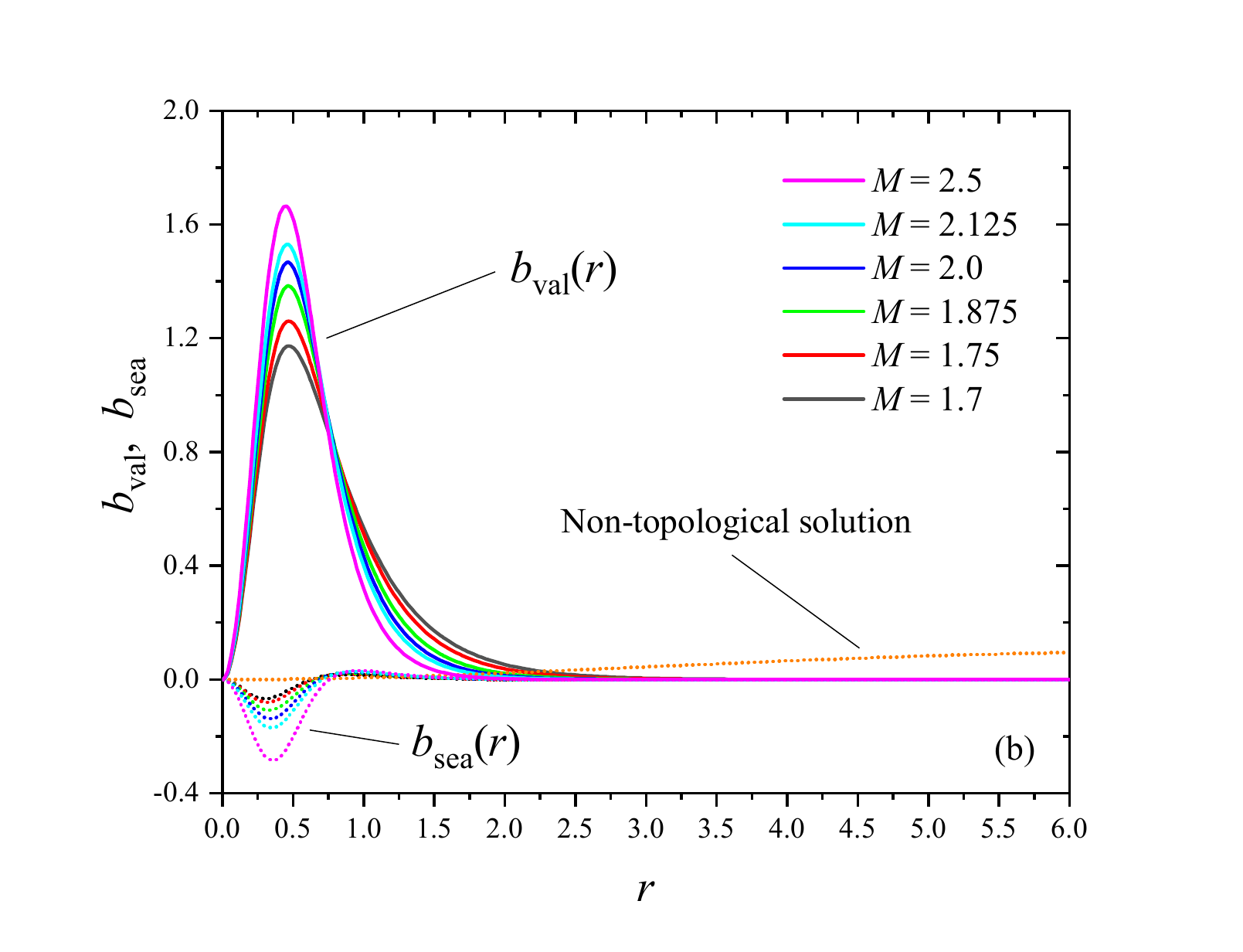}
	\\
	\includegraphics[width=0.5\linewidth]{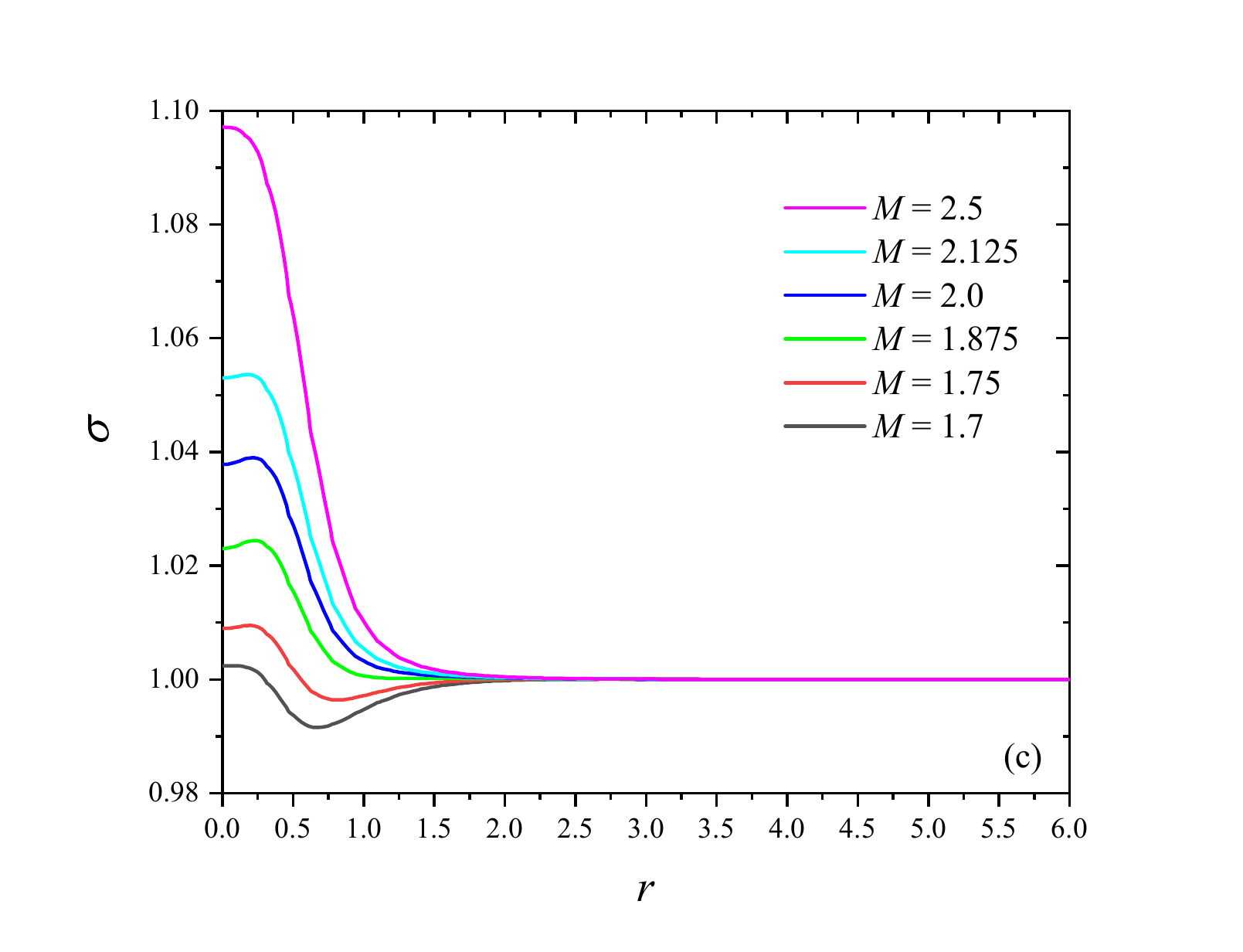}\hspace{-1.0cm}
	\includegraphics[width=0.5\linewidth]{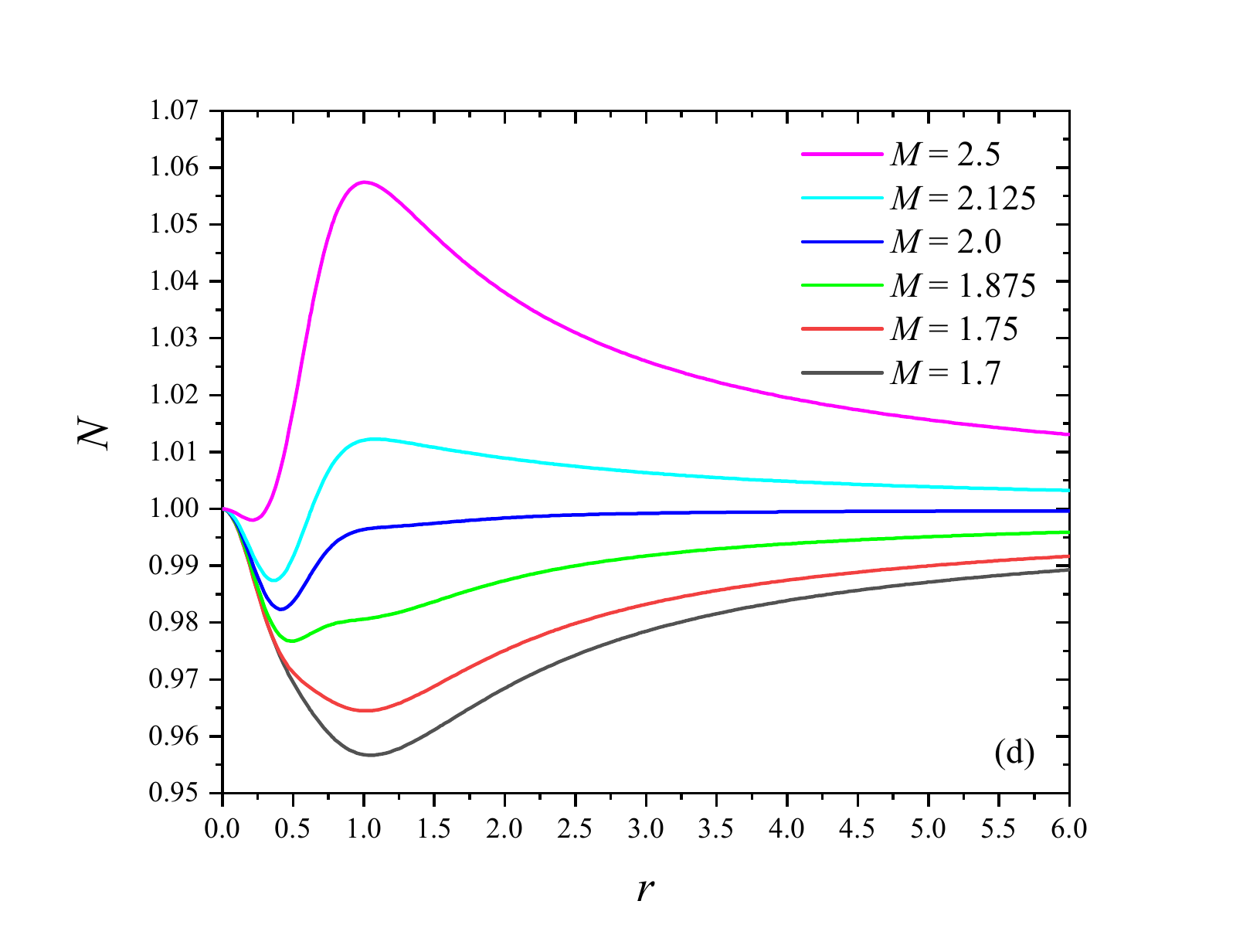}
	
	\caption{\label{Solution01}The self-consistent solutions of $\alpha^2=0.1$ 
	for several dynamical mass $M=1.7, 1.75, 1.875, 2.0, 2.125, 2.5$. 
	(a) the profile function $F(r)$, (b) the fermion density.  
	(c) the metric function $\sigma(r)$, and (d) the metric function $N(r)$. We also plot the profile function 
	and the fermion density of the non-topological solution of $M=1.675$, which will be discussed in Sec.IVB. 
	}
\end{figure*}


\begin{figure}[t]
	\centering 
	\includegraphics[width=1.0\linewidth]{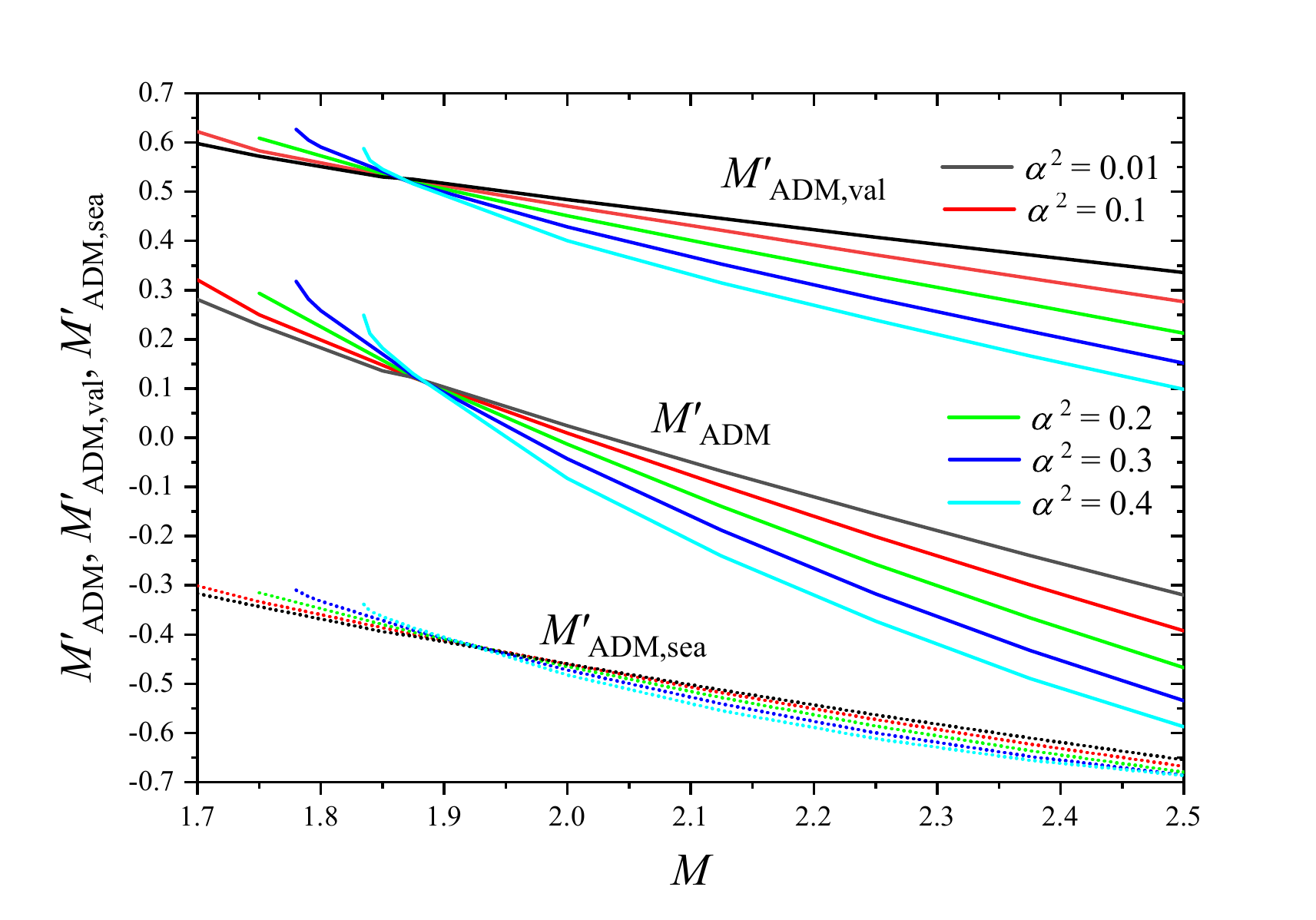}

	\caption{\label{ADM01}The valence, the Dirac-sea and their sum of the ADM mass defined by \eqref{ADMmass} 
	with the dynamical mass $M:[1.7,2.5]$ for the several gravitational 
	coupling constant $\alpha^2=0.01,0.1,0.2,0.3,0.4$ is plotted. 
	When $M$ increases, the ADM mass decreases and at some critical $M$, it becomes negative. 
	}
\end{figure}

\begin{figure*}[htbp]
	\centering 
	\includegraphics[width=0.5\linewidth]{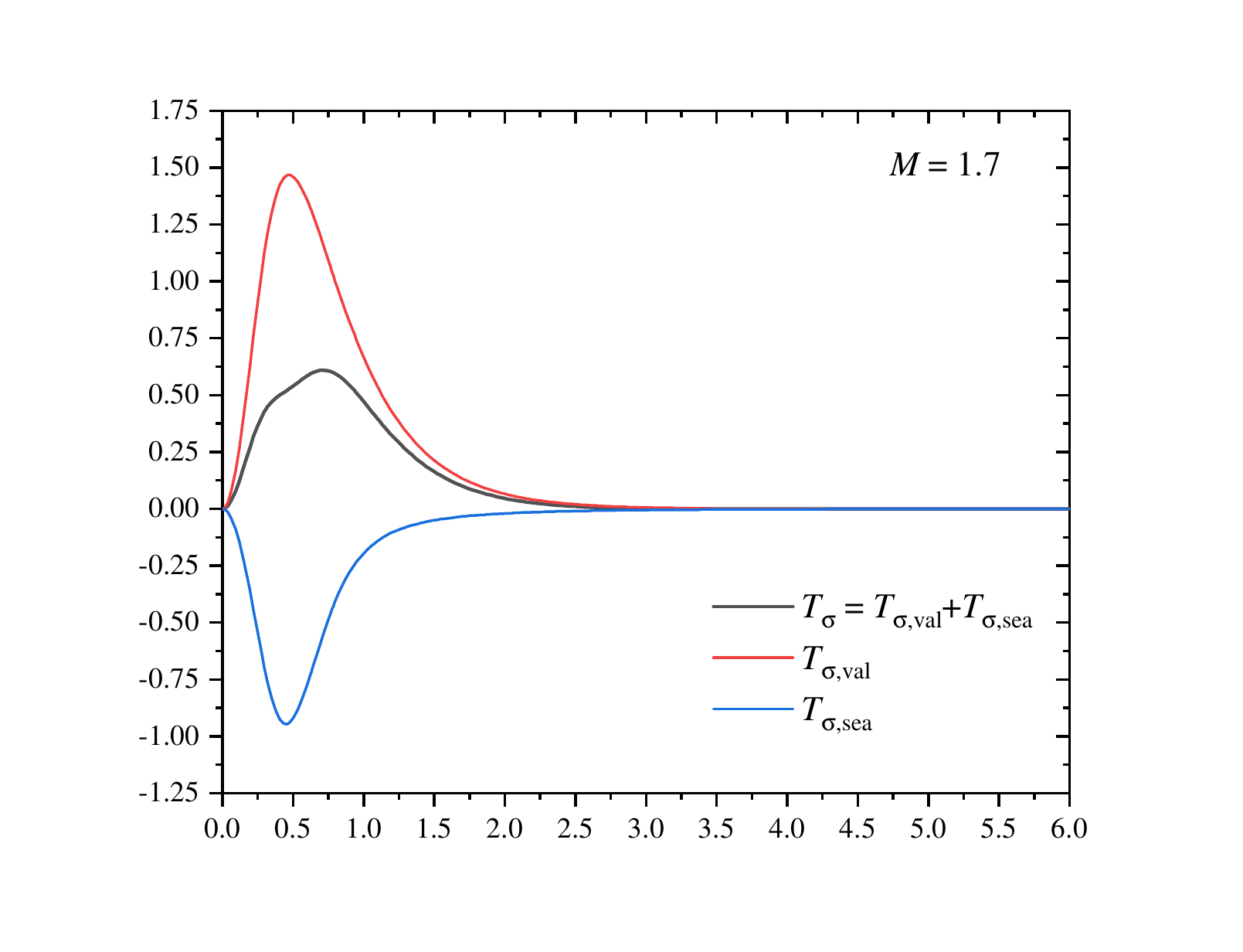}\hspace{-2.0cm}
	\includegraphics[width=0.5\linewidth]{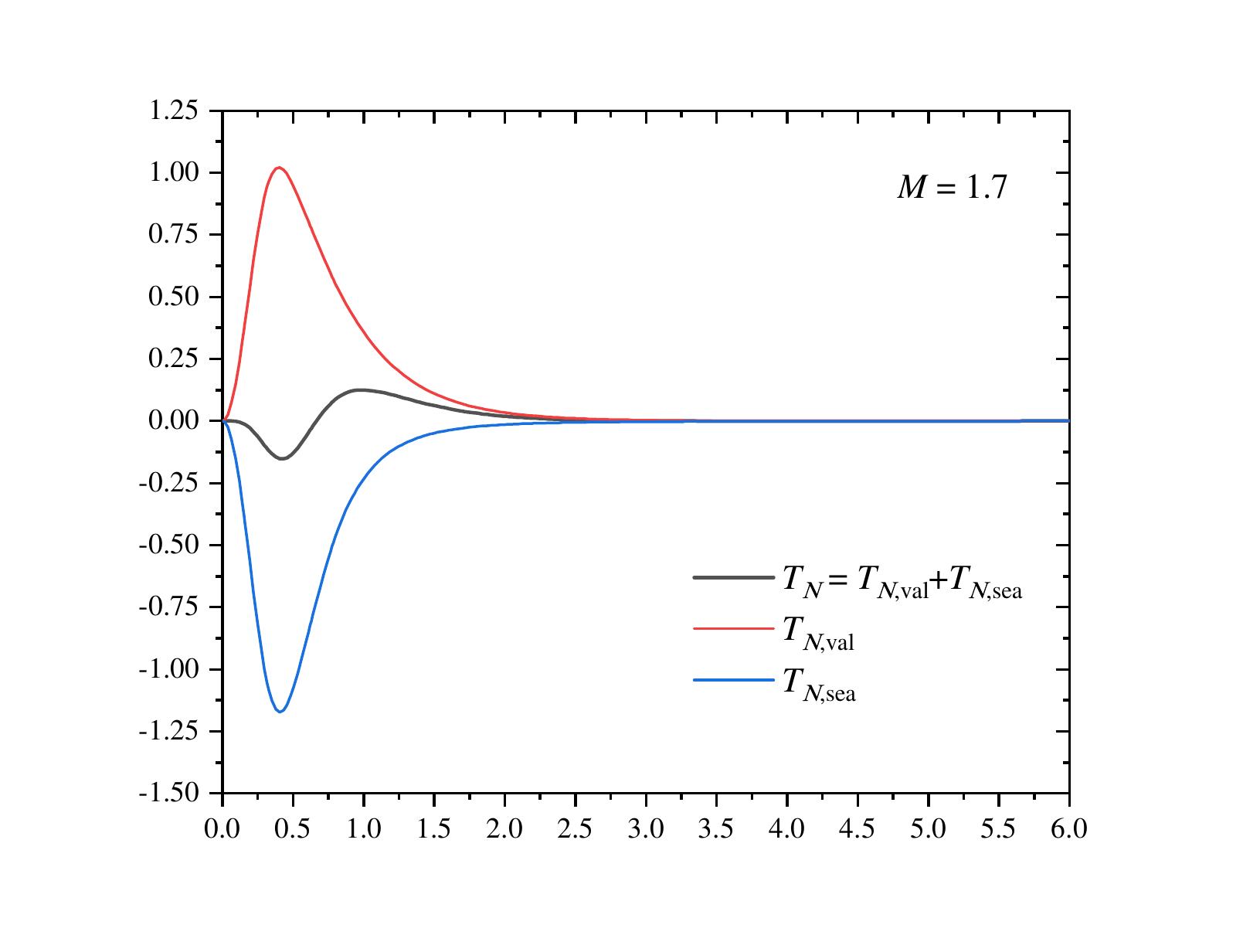}
	\\
	\vspace{-1cm}
	
	\includegraphics[width=0.5\linewidth]{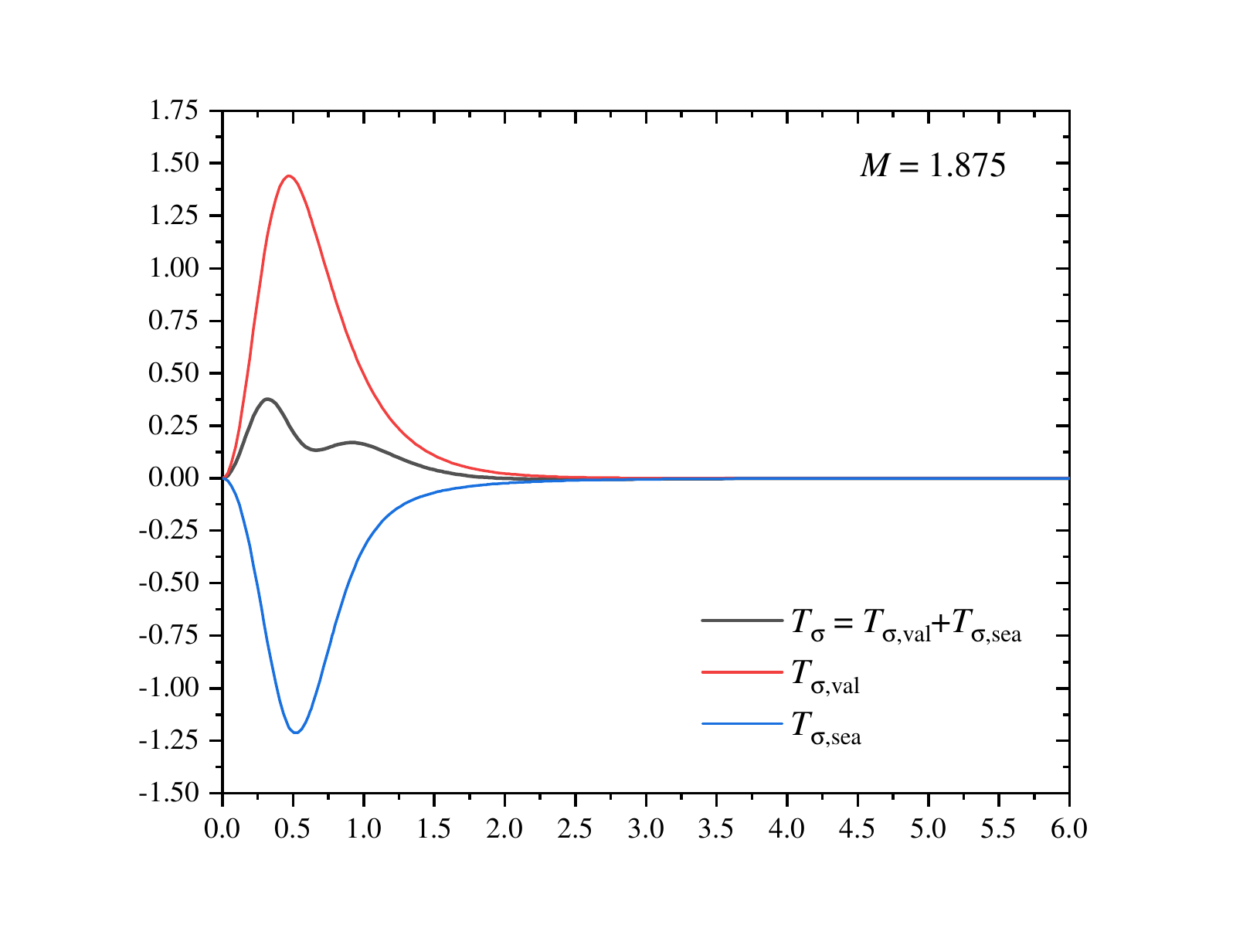}\hspace{-2.0cm}
	\includegraphics[width=0.5\linewidth]{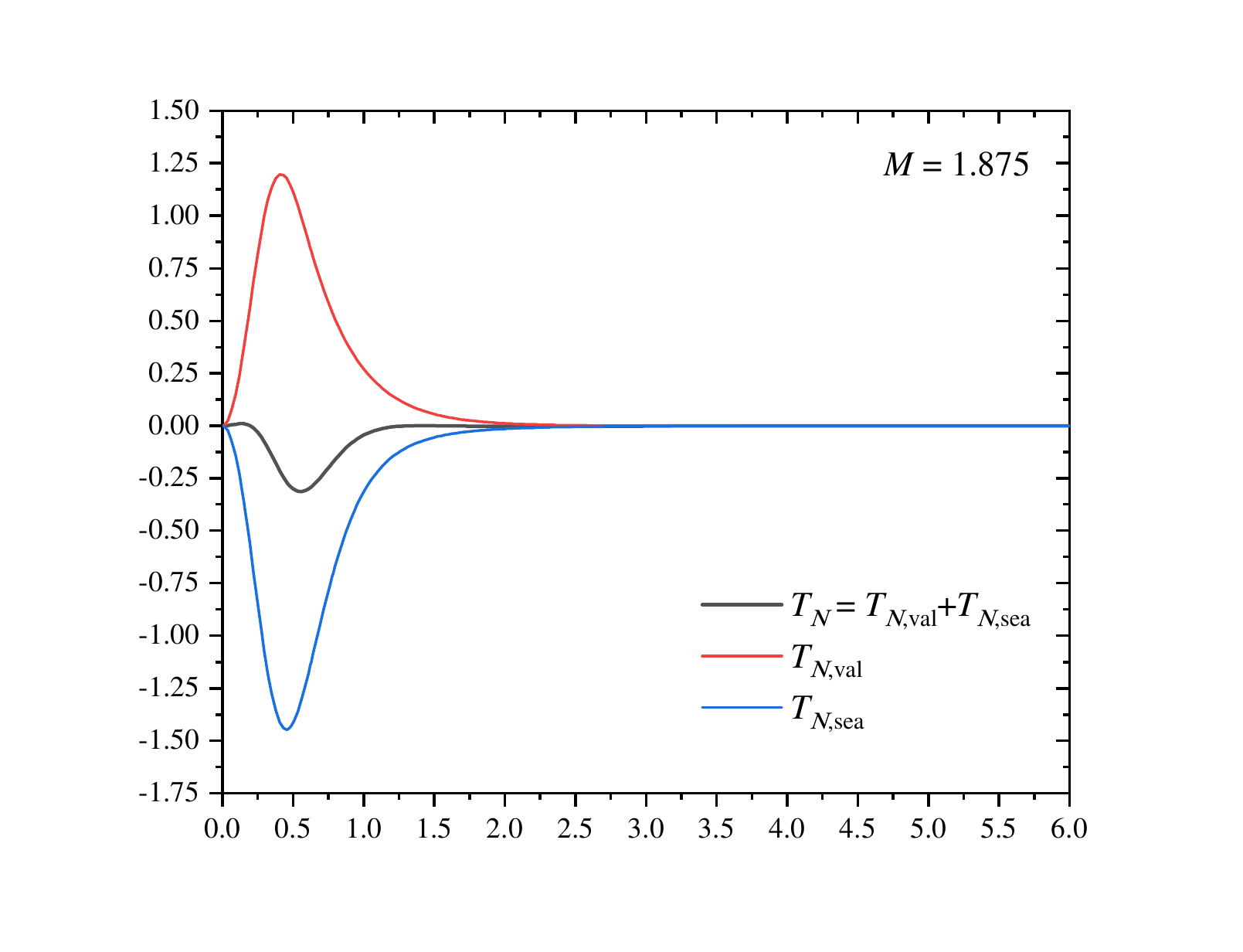}
	\\
	\vspace{-1cm}
	
	\includegraphics[width=0.5\linewidth]{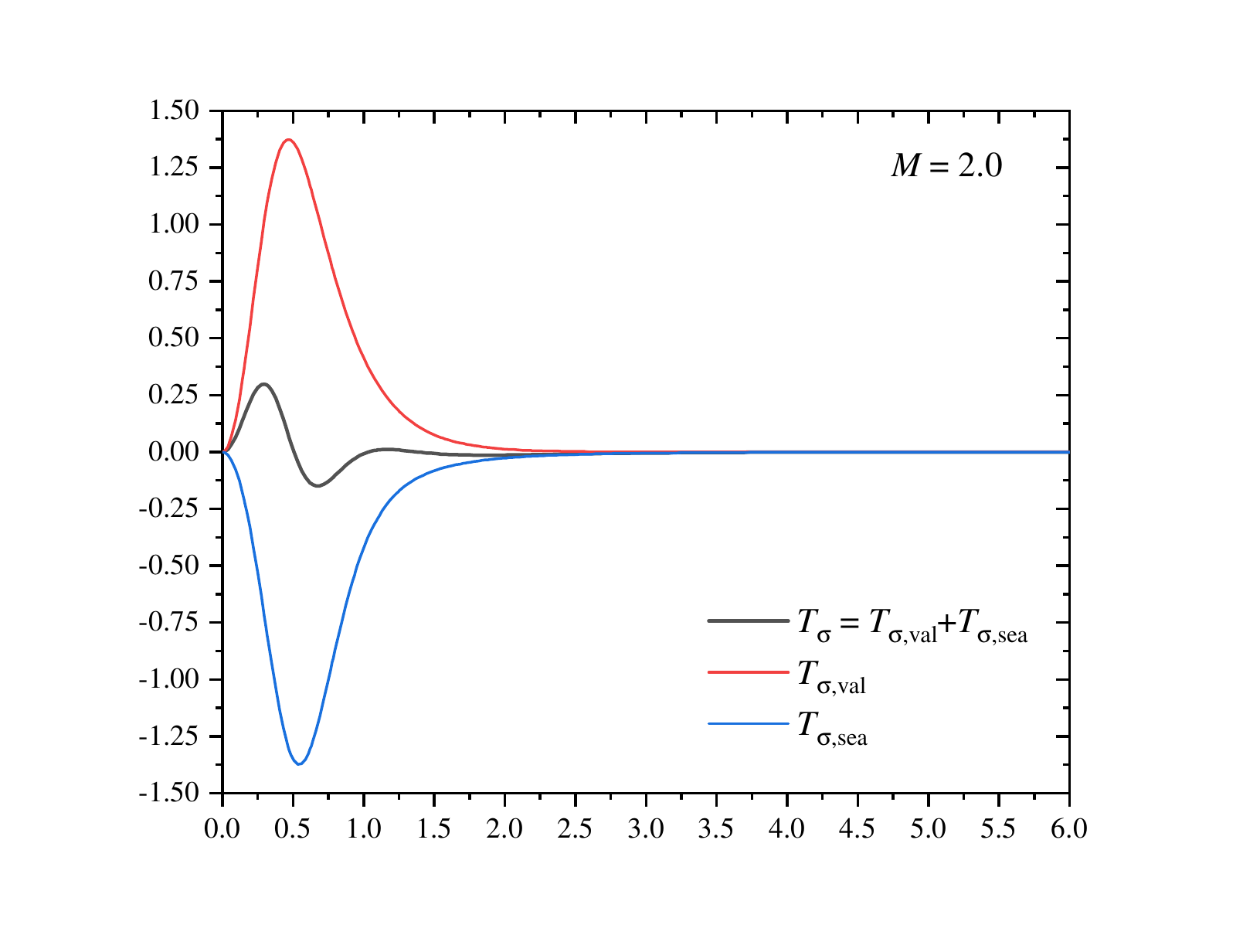}\hspace{-2.0cm}
	\includegraphics[width=0.5\linewidth]{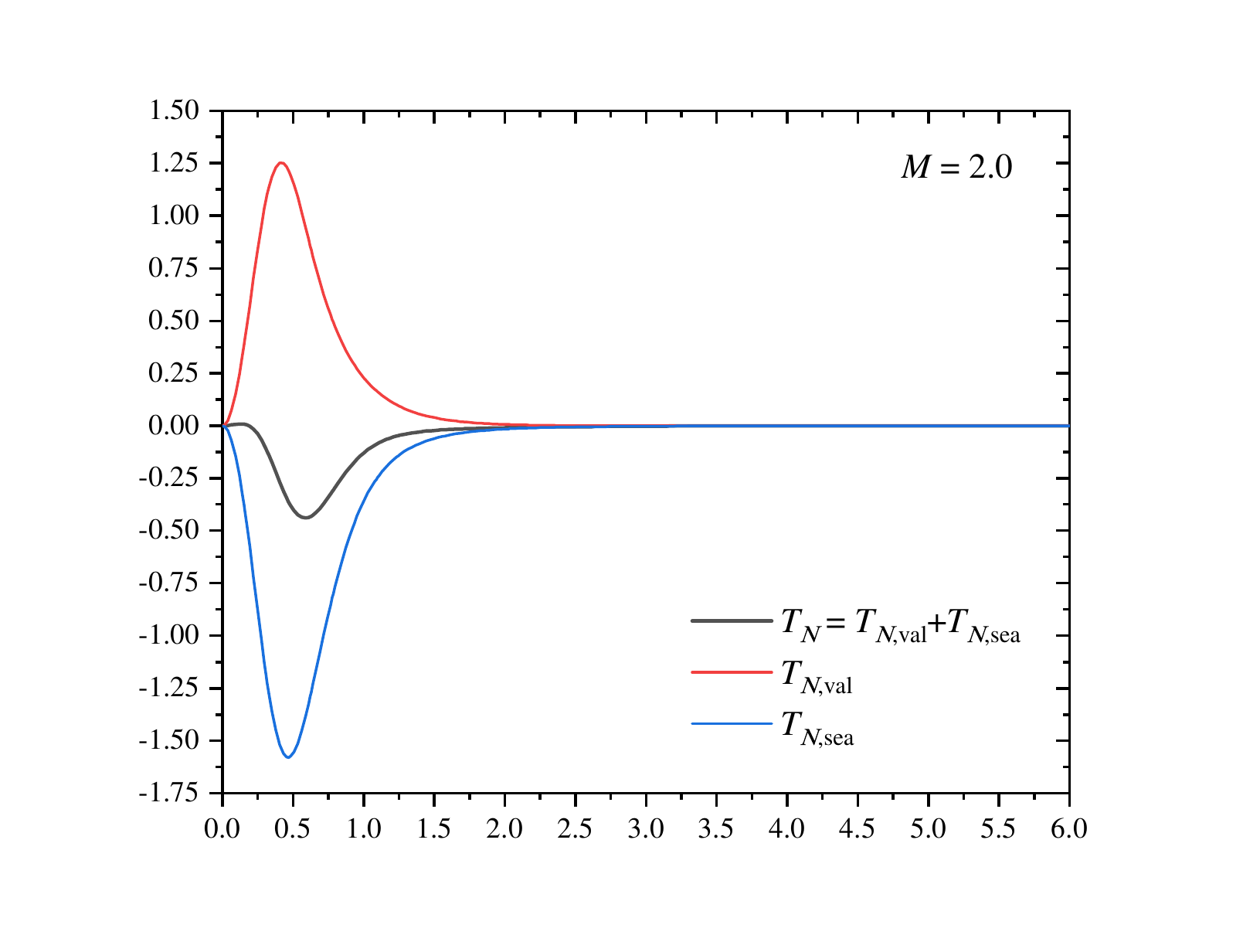}
	\\
	\vspace{-1cm}
	
	\includegraphics[width=0.5\linewidth]{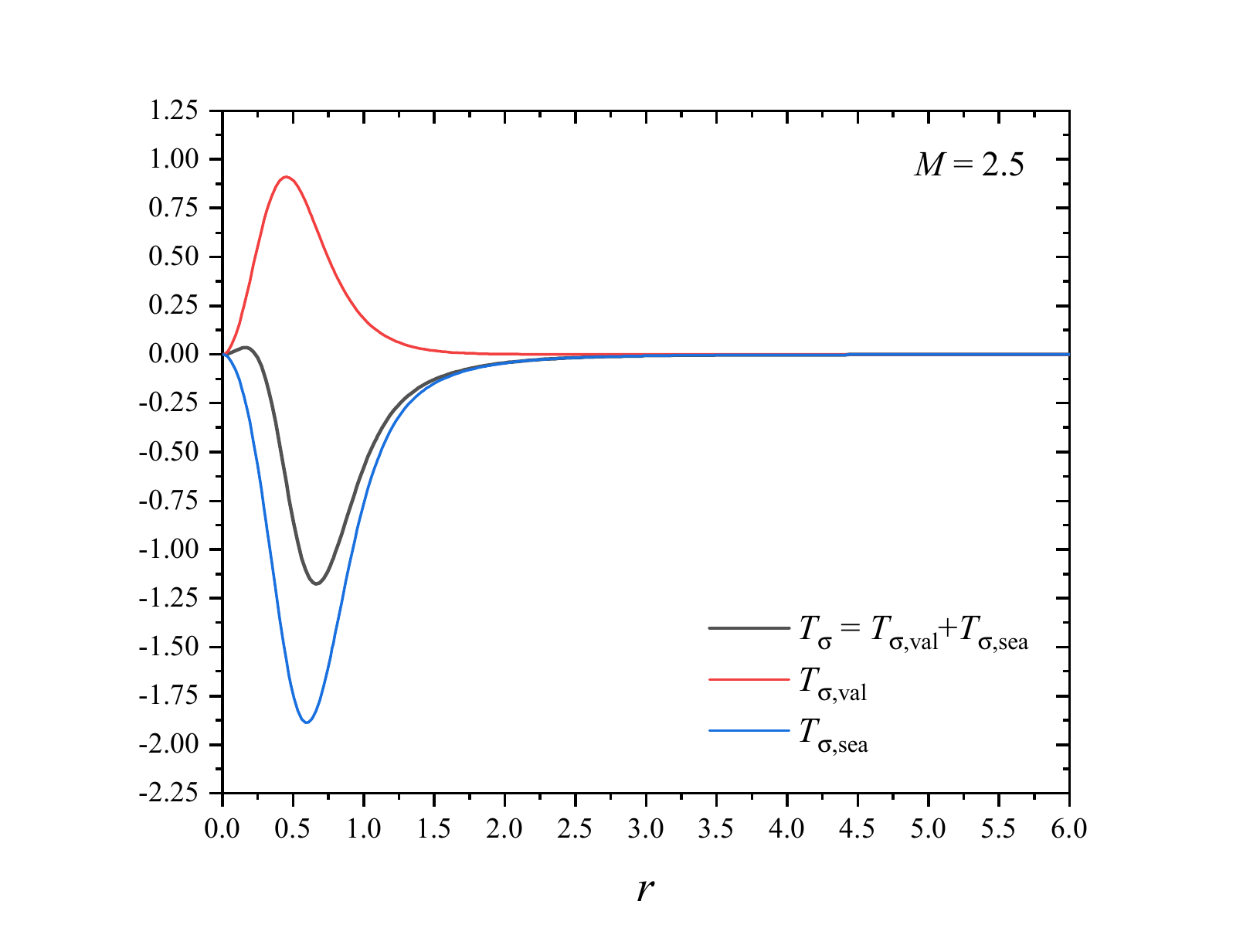}\hspace{-2.0cm}
	\includegraphics[width=0.5\linewidth]{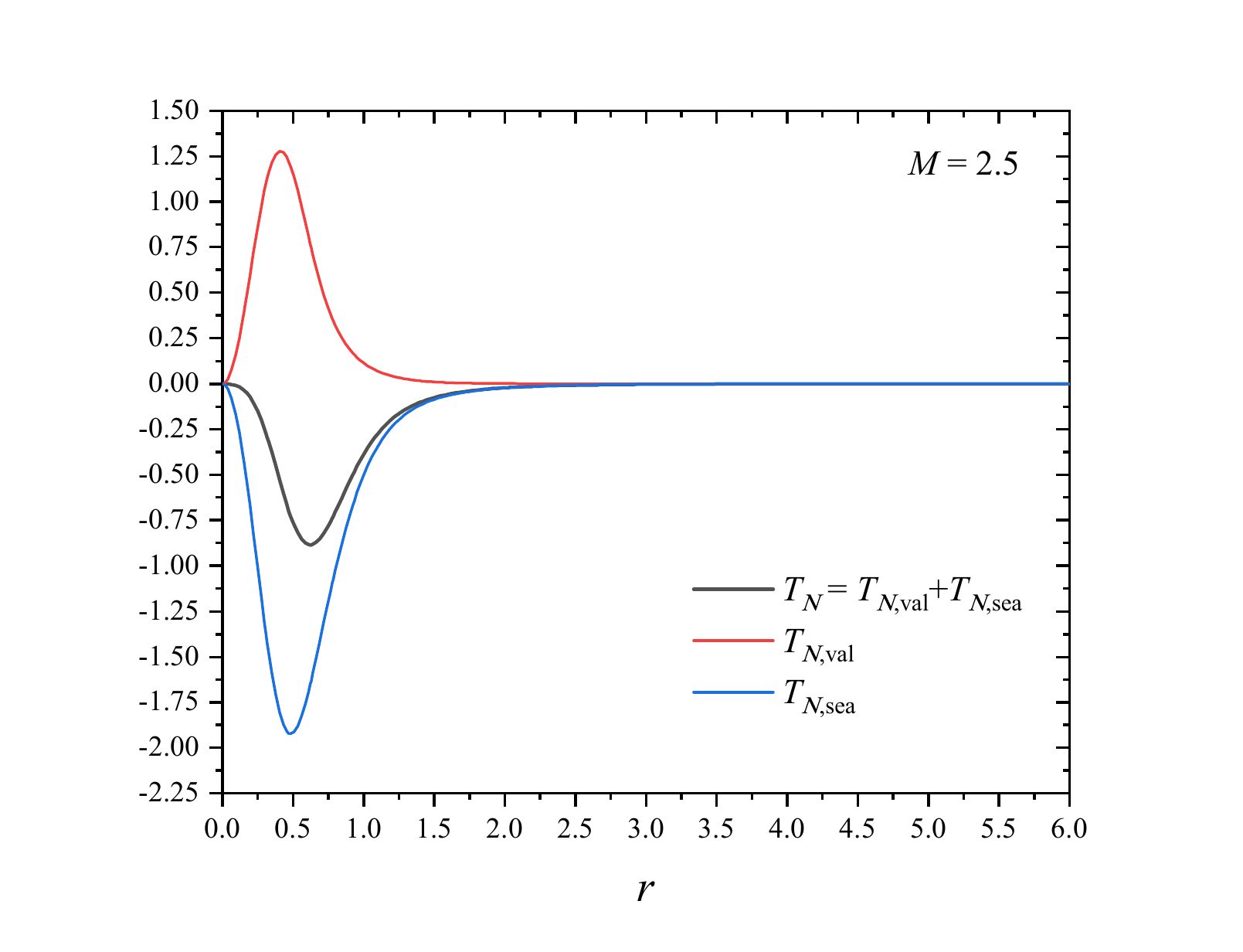}
	\\
	\caption{\label{EMtensor01}The source terms
	\eqref{EMtensors2},\eqref{EMtensorN2} 
	for the several dynamical mass 
	$M=1.7, 1.875, 2.0$ and $2.5$. $\alpha^2=0.1$.}
\end{figure*}

\begin{figure*}[htbp]
	\centering 
	\includegraphics[width=0.5\linewidth]{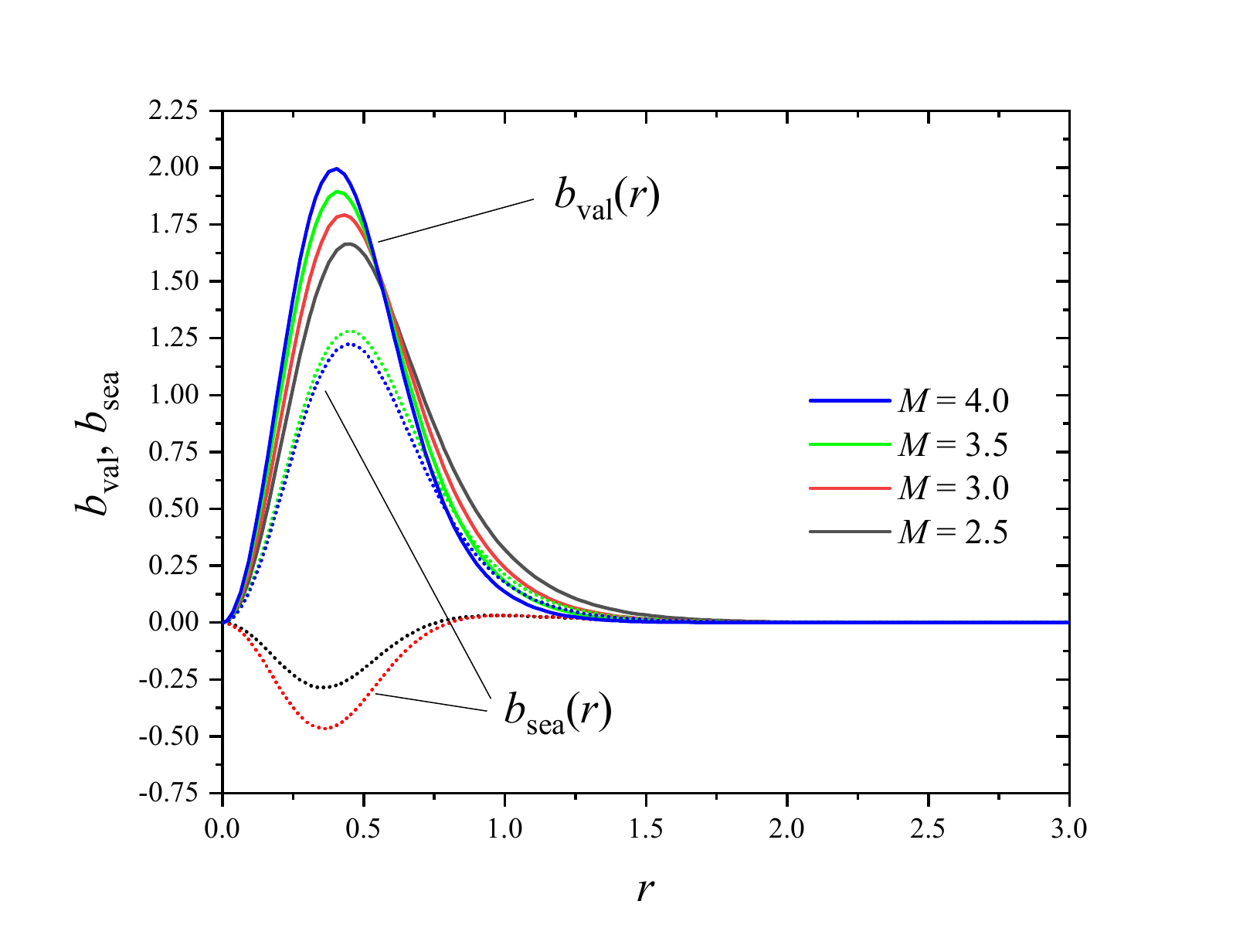}\hspace{-1.0cm}
	\includegraphics[width=0.5\linewidth]{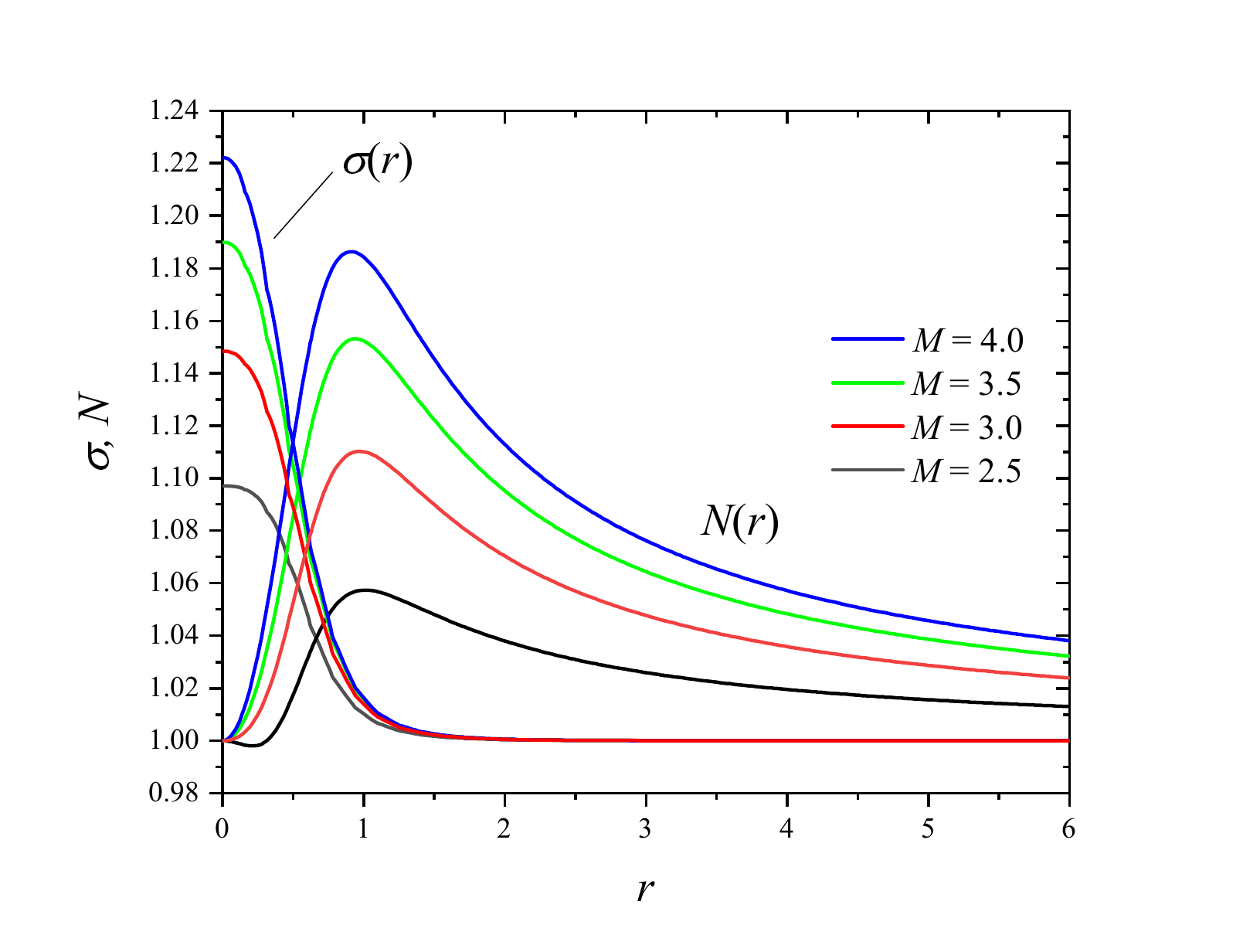}
	
	\caption{\label{Solution_high}The fermion density and the metric functions of $\alpha^2=0.1$ for larger 
	dynamical mass $M$. In these cases, it is evident that the ADM mass is negative.
	}
\end{figure*}


\subsection{Eigenproblem of the Dirac equation}

To numerically compute the Dirac equations \eqref{DiracEigen} and \eqref{DiracEigen0}, 
we employ the well-known matrix diagonalization scheme 
in terms of the Kahana-Ripka (KR) basis~\cite{Kahana:1984be}, 
which has particularly been extensively used in chiral invariant models coupled with 
the skyrmions. 
Note stress that the KR basis satisfy the normal orthogonality, such as
\begin{align}
\langle u_\mu | u_\nu \rangle :=\int d^3x u_\mu^\dagger (\bm{x})u_\nu(\bm{x}) =\delta_{\mu\nu}\,.
\end{align}
Comparing it with \eqref{orthogonality}, 
we can define auxiliary spinor functions $u_\mu(\bm{x})$ as~\cite{Mukhanov_Winitzki_2007}
\begin{align}
u_\mu(\bm{x}):=N^{-1/4}\phi_\mu(\bm{x})
\end{align}
which satisfies 
$\langle u_\mu |u_\nu \rangle \equiv (\phi_\mu,\phi_\nu)=\delta_{\mu\nu}$.
The eigenequation \eqref{DiracEigen} for $u_\mu$ becomes
\begin{align}
\hat{H}u_\mu =\epsilon_\mu u_\mu\,.
\label{peigenstates}
\end{align}
The pseudo-Hamiltonian $\tilde{H}$ is defined as 
\begin{align}
&\hat{H}:=N^{-1/4}\tilde{H}N^{1/4}
\nonumber \\
&\hspace{4mm}=(g^{tt})^{-1}\underline{\gamma^t}[-i(\underline{\gamma}^r\mathcal{\tilde{D}}_r
+\underline{\gamma}^\theta\partial_\theta+\underline{\gamma}^\phi\partial_\phi)
+MU^{\underline{\gamma}_5}]\,,
\nonumber \\
&\mathcal{\tilde{D}}_r:=\partial_r+\frac{1}{r}\biggl(1-\frac{1}{\sqrt{N}}\biggr)
+\frac{1}{2}(\log\sigma N)'\,.
\label{pseudoH}
\end{align}

It is helpful to introduce a plain-wave basis known as the KR basis~
in order to solve the Dirac eigenproblem \eqref{peigenstates} with the huge 
number of positive and negative eigenstates simultaneously. The method commonly has used 
in many literatures~\cite{Diakonov:1987ty, Wakamatsu:1990ud, Sawado:2004pm}. 
The Hamiltonian with hedgehog ansatz \eqref{pseudoH} commutes with the parity and the grandspin operator 
given by  
\begin{align}
	\bm{K}=\bm{j}+\bm{\tau}/2=\bm{l}+\bm{\sigma}/2+\bm{\tau}/2,
\end{align}
where $\bm{j},\bm{l}$ are respectively total angular momentum and orbital angular momentum.  
Accordingly, the angular basis can be written as  
\begin{align}
|(lj)KM\rangle= \sum_{j_3\tau_3}C^{KM}_{jj_3\frac{1}{2}\tau_3}
\Bigl(\sum_{m\sigma_3}C^{jj_3}_{lm\frac{1}{2}\sigma_3}
|lm \rangle |\frac{1}{2}\sigma_3 \rangle \Bigr) |\frac{1}{2} \tau_3 \rangle\,,\nonumber
\end{align}
and the following states are possible 
\begin{align}
&|0\rangle =|(K~K+\frac{1}{2})KM \rangle\,,~~  
|1\rangle =|(K~K-\frac{1}{2})KM \rangle\,,  \nonumber   \\
&|2\rangle =|(K+1 K+\frac{1}{2})KM\rangle\,,~~
|3\rangle =|(K-1 K-\frac{1}{2})KM\rangle\,. \nonumber
\end{align}
It is useful to examine the matrix element of $\bm{\tau}\cdot\hat{\bm{r}}$ in terms of these basis
which are Hermite and 
\begin{align}
&\langle 0|\bm{\tau}\cdot\hat{\bm{r}}|2\rangle = \frac{1}{2K+1},~~
\langle 0|\bm{\tau}\cdot\hat{\bm{r}}|3\rangle = \frac{-2\sqrt{K(K+1)}}{2K+1},
\nonumber \\
&\langle 1|\bm{\tau}\cdot\hat{\bm{r}}|2\rangle=\langle 0|\bm{\tau}\cdot\hat{\bm{r}}|3\rangle,~~
\langle 1|\bm{\tau}\cdot\hat{\bm{r}}|3\rangle=-\langle 0|\bm{\tau}\cdot\hat{\bm{r}}|2\rangle
\end{align}
and zero for all others. 
With this angular basis, the normalized eigenstates of the free Hamiltonian 
in a spherical box with radius $R$ can be constructed as follows:
The states with $K^\mathcal{P}=0^+,1^-,2^+,\cdots$ of the 
parity $\mathcal{P}$ has four plain-wave bases as we call \textit{the natural bases}
\begin{align}
&\langle\bm{r}|u^{(a)}_{KM}\rangle_\textrm{n} =
N_k\left( 
\begin{array}{c}
i\omega^{+}_{k,\epsilon}j_{K}(kr)|0\rangle \\
\omega^{-}_{k,\epsilon}j_{K+1}(kr)|2\rangle
\end{array}
\right), \nonumber \\
&\langle\bm{r}|u^{(b)}_{KM}\rangle_\textrm{n} =
N_k\left( 
\begin{array}{c}
i\omega^{+}_{k,\epsilon}j_{K}(kr)|1\rangle \\
-\omega^{-}_{k,\epsilon}j_{K-1}(kr)|3\rangle
\end{array}
\right), 
\end{align}
and the another set of the bases with $K^\mathcal{P}=0^-,1^+,2^-,\cdots$ (\textit{the unnatural bases}) 
are
\begin{align}
&\langle\bm{r}|u^{(a)}_{KM}\rangle_\textrm{u} =
N_k\left( 
\begin{array}{c}
i\omega^{+}_{k,\epsilon}j_{K+1}(kr)|2\rangle \\
-\omega^{-}_{k,\epsilon} j_{K}(kr)|0\rangle
\end{array}
\right), \nonumber \\
&\langle\bm{r}|u^{(b)}_{KM}\rangle_\textrm{u} =
N_k\left( 
\begin{array}{c}
i\omega^{+}_{k,\epsilon}j_{K-1}(kr)|3\rangle \\
\omega^{-}_{k,\epsilon} j_{K}(kr)|1\rangle
\end{array}
\right). 
\label{kahana_ripka}
\end{align}
The normalization constant
	$N_k=[\frac{1}{2}R^3(j_{K+1}(kR))^2]^{-1/2}$ 
and also, 
there are two types of coefficients which corresponds to the sign of the 
energy eigenvalues $\epsilon$,
$\omega^{+}_{k,\epsilon>0},\omega^{-}_{k,\epsilon<0}={\rm sgn}(\epsilon_k), 
\omega^{-}_{k,\epsilon>0},\omega^{+}_{k,\epsilon<0}=k/(\epsilon_k+M)$.
The momenta are discretized by the boundary condition
\begin{align}
j_K(k_i R)=0.
\end{align} 
The orthogonality of the basis is then satisfied by  
\begin{align}
&\int^R_0 dr r^2 j_K(k_i r)j_K(k_j r)
=\int^R_0 dr r^2 j_{K\pm 1}(k_i r)j_{K\pm 1}(k_j r)  \nonumber \\
&=\delta_{ij}\frac{R^3}{2}  [j_{K\pm 1}(k_i R)]^2 \, .
\label{orthogonality_KH}
\end{align}
We truncate the size of our setup to easily carry out our computation and 
set the maximum values of $K_\textrm{max}, k_\textrm{max}$ as
\begin{align}
K_\textrm{max}=16,~~k_\textrm{max}=128,~R=20.0, 
\nonumber 
\end{align}
which realizes the sufficient numerical convergence. 

We construct the matrix form of the effective Hamiltonian in terms of the natural and the unnatural bases as
\begin{align}
_\textrm{n}\langle u_{KM}|\hat{H}| u_{KM}\rangle_\textrm{n},~~_\textrm{u}\langle u_{KM}|\hat{H}| u_{KM}\rangle_\textrm{u},.
\end{align}
which are diagonalized numerically with each $K$. 
Since we will need them later to estimate the source term (the stress-energy tensors) of the Einstein equations, 
it is worthwhile to present explicit form of some components of the effective Hamiltonian (though they are a little complicated).
Some useful formalism for the explicit evaluation of the matrix elements are summarized in Appendix.

In terms of the natural bases, the Hermitian form of the matrix elements are given by
\begin{widetext}
\begin{align}
&\frac{1}{2}[{}_\textrm{n}\langle u^{(a)}_{KM}|Hu^{(a)}_{KM}\rangle_\textrm{n}
-{}_\textrm{n}\langle Hu^{(a)}_{KM}|u^{(a)}_{KM}\rangle_\textrm{n}]
\nonumber \\
&=2\pi N_kN_{k'}\int r^2dr \sigma\sqrt{N}
\biggl\{(1-\sqrt{N})\Bigl(\frac{K+1}{r}\omega^+_{k,\epsilon}j_K(kr)j_{K+1}(k'r)\omega^-_{k',\epsilon'}
+\frac{K+1}{r}\omega^-_{k,\epsilon}j_{K+1}(kr)j_K(k'r)\omega^+_{k',\epsilon'}\Bigr)
\nonumber \\
&+k'\sqrt{N}\Bigl(\omega^+_{k,\epsilon}j_K(kr)j_K(k'r)\omega^-_{k',\epsilon'}
+\omega^-_{k,\epsilon}j_{K+1}(kr)j_{K+1}(k'r)\omega^+_{k',\epsilon'}\Bigr)
+(k,\epsilon) \leftrightarrow (k',\epsilon')\biggr\}
\nonumber \\
&+4\pi N_kN_{k'}\int r^2dr \sigma\sqrt{N}
\biggl\{M\cos F\Bigl(\omega^+_{k,\epsilon}j_K(kr)j_K(k'r)\omega^+_{k',\epsilon'}
-\omega^-_{k,\epsilon}j_{K+1}(kr)j_{K+1}(k'r)\omega^-_{k',\epsilon'}\Bigr)
\nonumber \\
&+M\sin F\Bigl(\omega^+_{k,\epsilon}j_K(kr)j_{K+1}(k'r)\omega^-_{k',\epsilon'}\langle 0|\bm{\tau}\cdot\hat{\bm{r}}|2\rangle
+\omega^-_{k,\epsilon}j_{K+1}(kr)j_{K}(k'r)\omega^+_{k',\epsilon'}\langle 2|\bm{\tau}\cdot\hat{\bm{r}}|0\rangle)\biggr\}\,,
\label{matnaa}
\end{align}
\begin{align}
&\frac{1}{2}[{}_\textrm{n}\langle u^{(a)}_{KM}|Hu^{(b)}_{KM}\rangle_\textrm{n}
-{}_\textrm{n}\langle Hu^{(a)}_{KM}|u^{(b)}_{KM}\rangle_\textrm{n}]
\nonumber \\
&=4\pi N_kN_{k'}\int r^2dr \sigma\sqrt{N}
\biggl\{M\sin F\Bigl(-\omega^+_{k,\epsilon}j_K(kr)j_{K-1}(k'r)\omega^-_{k',\epsilon'}\langle 0|\bm{\tau}\cdot\hat{\bm{r}}|3\rangle
+\omega^-_{k,\epsilon}j_{K+1}(kr)j_{K}(k'r)\omega^-_{k',\epsilon'}\langle 2|\bm{\tau}\cdot\hat{\bm{r}}|1\rangle)\biggr\}\,,
\label{matnab}
\end{align}
\begin{align}
&\frac{1}{2}[{}_\textrm{n}\langle u^{(b)}_{KM}|Hu^{(b)}_{KM}\rangle_\textrm{n}
-{}_\textrm{n}\langle Hu^{(b)}_{KM}|u^{(b)}_{KM}\rangle_\textrm{n}]
\nonumber \\
&=2\pi N_kN_{k'}\int r^2dr \sigma\sqrt{N}
\biggl\{-(1-\sqrt{N})\Bigl(\frac{K-2}{r}\omega^+_{k,\epsilon}j_K(kr)j_{K-1}(k'r)\omega^-_{k',\epsilon'}
+\frac{K}{r}\omega^-_{k,\epsilon}j_{K-1}(kr)j_K(k'r)\omega^+_{k',\epsilon'}\Bigr)
\nonumber \\
&-k'\sqrt{N}\Bigl(\omega^+_{k,\epsilon}j_K(kr)j_K(k'r)\omega^-_{k',\epsilon'}
+\omega^-_{k,\epsilon}jk_{K-1}(kr)j_{K-1}(k'r)\omega^+_{k',\epsilon'}\Bigr)
+(k,\epsilon) \leftrightarrow (k',\epsilon')\biggr\}
\nonumber \\
&+4\pi N_kN_{k'}\int r^2dr \sigma\sqrt{N}
\biggl\{M\cos F\Bigl(\omega^+_{k,\epsilon}j_K(kr)j_K(k'r)\omega^+_{k',\epsilon'}
-\omega^-_{k.\epsilon}j_{K+1}(kr)j_{K+1}(k'r)\omega^-_{k',\epsilon'}\Bigr)
\nonumber \\
&-M\sin F\Bigl(\omega^+_{k,\epsilon}j_K(kr)j_{K-1}(k'r)\omega^-_{k',\epsilon'}\langle 1|\bm{\tau}\cdot\hat{\bm{r}}|3\rangle
+\omega^-_{k,\epsilon}j_{K-1}(kr)j_{K}(k'r)\omega^+_{k',\epsilon'}\langle 3|\bm{\tau}\cdot\hat{\bm{r}}|1\rangle)\biggr\}\,,.
\label{matnbb}
\end{align}
Also, the matrix elements of the unnatural bases are
\begin{align}
&\frac{1}{2}[{}_\textrm{u}\langle u^{(a)}_{KM}|Hv^{(a)}_{KM}\rangle_\textrm{u}
-{}_\textrm{u}\langle Hv^{(a)}_{KM}|u^{(a)}_{KM}\rangle_\textrm{u}]
\nonumber \\
&=2\pi N_kN_{k'}\int r^2dr \sigma\sqrt{N}
\biggl\{(1-\sqrt{N})\frac{K+1}{r}\Bigl(\omega^+_{k,\epsilon}j_{K+1}(kr)j_{K}(k'r)\omega^-_{k',\epsilon'}
+\omega^-_{k,\epsilon}j_{K}(kr)j_{K+1}(k'r)\omega^+_{k',\epsilon'}\Bigr)
\nonumber \\
&+k'\sqrt{N}\Bigl(\omega^+_{k,\epsilon}j_{K+1}(kr)j_{K+1}(k'r)\omega^-_{k',\epsilon'}
+\omega^-_{k,\epsilon}j_{K}(kr)j_{K}(k'r)\omega^+_{k',\epsilon'}\Bigr)
+(k,\epsilon) \leftrightarrow (k',\epsilon')\biggr\}
\nonumber \\
&+4\pi N_kN_{k'}\int r^2dr \sigma\sqrt{N}
\biggl\{M\cos F\Bigl(\omega^+_{k,\epsilon}j_{K+1}(kr)j_{K+1}(k'r)\omega^+_{k',\epsilon'}
-\omega^-_{k,\epsilon}j_{K}(kr)j_{K}(k'r)\omega^-_{k',\epsilon'}\Bigr)
\nonumber \\
&-M\sin F\Bigl(\omega^+_{k,\epsilon}j_{K+1}(kr)j_{K}(k'r)\omega^-_{k',\epsilon'}\langle 2|\bm{\tau}\cdot\hat{\bm{r}}|0\rangle
+\omega^-_{k,\epsilon}j_{K}(kr)j_{K+1}(k'r)\omega^+_{k',\epsilon'}\langle 0|\bm{\tau}\cdot\hat{\bm{r}}|2\rangle)\biggr\}\,,
\label{matuaa}
\end{align}
\begin{align}
&\frac{1}{2}[{}_\textrm{u}\langle u^{(a)}_{KM}|Hv^{(b)}_{KM}\rangle_\textrm{u}
-{}_\textrm{u}\langle Hv^{(a)}_{KM}|u^{(b)}_{KM}\rangle_\textrm{u}]
\nonumber \\
&=4\pi N_kN_{k'}\int r^2dr \sigma\sqrt{N}
\biggl\{M\sin F\Bigl(\omega^+_{k,\epsilon}j_{K+1}(kr)j_{K}(k'r)\omega^-_{k',\epsilon'}\langle 2|\bm{\tau}\cdot\hat{\bm{r}}|1\rangle
-\omega^-_{k,\epsilon}j_{K}(kr)j_{K-1}(k'r)\omega^-_{k',\epsilon'}\langle 0|\bm{\tau}\cdot\hat{\bm{r}}|3\rangle)\biggr\}\,,
\label{matuab}
\end{align}
\begin{align}
&\frac{1}{2}[{}_\textrm{u}\langle u^{(b)}_{KM}|Hv^{(b)}_{KM}\rangle_\textrm{u}
-{}_\textrm{u}\langle Hv^{(b)}_{KM}|u^{(b)}_{KM}\rangle_\textrm{u}]
\nonumber \\
&=2\pi N_kN_{k'}\int r^2dr \sigma\sqrt{N}
\biggl\{(1-\sqrt{N})\Bigl(\frac{K}{r}\omega^+_{k,\epsilon}j_{K}(kr)j_{K-1}(k'r)\omega^-_{k',\epsilon'}
+\frac{K-2}{r}\omega^-_{k,\epsilon}j_{K-1}(kr)j_K(k'r)\omega^+_{k',\epsilon'}\Bigr)
\nonumber \\
&+k'\sqrt{N}\Bigl(\omega^+_{k,\epsilon}j_K(kr)j_K(k'r)\omega^-_{k',\epsilon'}
+\omega^-_{k,\epsilon}jk_{K-1}(kr)j_{K-1}(k'r)\omega^+_{k',\epsilon'}\Bigr)
+(k,\epsilon) \leftrightarrow (k',\epsilon')\biggr\}
\nonumber \\
&+4\pi N_kN_{k'}\int r^2dr \sigma\sqrt{N}
\biggl\{M\cos F\Bigl(\omega^+_{k,\epsilon}j_K(kr)j_K(k'r)\omega^+_{k',\epsilon'}
-\omega^-_{k.\epsilon}j_{K+1}(kr)j_{K+1}(k'r)\omega^-_{k',\epsilon}\Bigr)
\nonumber \\
&+M\sin F\Bigl(\omega^+_{k,\epsilon}j_K(kr)j_{K-1}(k'r)\omega^-_{k',\epsilon'}\langle 3|\bm{\tau}\cdot\hat{\bm{r}}|1\rangle
+\omega^-_{k,\epsilon}j_{K-1}(kr)j_{K}(k'r)\omega^+_{k',\epsilon'}\langle 1|\bm{\tau}\cdot\hat{\bm{r}}|3\rangle)\biggr\}\,,
\label{matubb}
\end{align}
\end{widetext}
Only the combinations of $\sigma\sqrt{N}$ or $\sigma N$ involve the metric functions in these matrix elements, 
which will be helpful for subsequently directly evaluating the source terms.

The energy of the Dirac fermion is defined by $E_\textrm{D}:=-S_\textrm{D}$ and 
with the one-particle valence (the level crossing) eigenenergy $\varepsilon_0$, we write
\begin{align}
&E_\textrm{D}:=\theta(\varepsilon_0)E_\textrm{val}+E_\textrm{sea}\,,
\nonumber \\
&E_\textrm{val}=N_\textrm{c}\varepsilon_0\,,
\nonumber \\
&E_\textrm{sea}=N_\textrm{c}\sum_\mu\biggl\{\mathcal{N}(\varepsilon_\mu)|\varepsilon_\mu|
+\frac{\Lambda}{\sqrt{4\pi}}\exp\biggl(\frac{\varepsilon_\mu}{\Lambda}\biggr)^2\biggr\}
\nonumber \\
&\hspace{0.8cm}-N_\textrm{c}\sum_\mu\biggl\{\mathcal{N}(\varepsilon^0_k)|\varepsilon^0_k|
+\frac{\Lambda}{\sqrt{4\pi}}\exp\biggl(\frac{\varepsilon^0_k}{\Lambda}\biggr)^2\biggr\}\,.
\label{fermionenergy}
\end{align}
Also, we define radial number density of the fermion with the valence fermion state $u_0$
\begin{align}
&b(r):=\theta(\varepsilon_0)b_\textrm{val}(r)+b_\textrm{sea}(r)\,,
\nonumber \\ 
&b_\textrm{val}(r)=\int d\Omega r^2u_0^\dagger(\bm{x}) u_0(\bm{x})\,,
\nonumber \\
&b_\textrm{sea}(r)=\int d\Omega r^2\biggl[
\sum_\mu \textrm{sgn}(\varepsilon_\mu)\mathcal{N}(\varepsilon_\mu)_\mu u_\mu^\dagger(\bm{x}) u_\mu(\bm{x})
\nonumber \\
&\hspace{2cm}-\sum_k \textrm{sgn}(\varepsilon^0_k)\mathcal{N}_k(\varepsilon^0_k) u_k^{0\dagger}(\bm{x}) u^0_k(\bm{x})
\biggr]\,,
\label{fermiondensity}
\end{align}
where $d\Omega:=\sin\theta d\theta d\varphi$.

\subsection{The derivation of the metric functions}

In this section, we provide a brief explanation of how to integrate the Einstein equations 
to obtain the metric functions~\eqref{Einsteins},\eqref{EinsteinN}. 
The eigenstates of Eq.\eqref{peigenstates} are derived by the diagonalization procedure of
the matrix elements~\eqref{matnaa}-\eqref{matnbb} and \eqref{matuaa}-\eqref{matubb} which 
depend only on combination of the metric functions $\sigma\sqrt{N}$ or $\sigma N$. 
After a lengthy calculation, we find that 
the source terms~\eqref{EMtensors2} and \eqref{EMtensorN2} have the following form 
\begin{align}
&T_\sigma[N;r] = g_1(r)\sqrt{N(r)}+g_2(r)N(r)\,,
\nonumber \\
&T_N[\sigma,N;r]=\sigma (r)\Bigl(g_1(r)\frac{1}{2\sqrt{N(r)}}+g_2(r)\Bigr) 
\label{SET}
\end{align}
where $g_i(r),i=1,2$ can be estimated via the eigenstates of the Dirac equation. 
We solve the Einstein equations self-consistently, i.e., 
we assume $T_\sigma,T_N$ as the background fields and analytically integrate the equations in order.
When we get the metric functions $\sigma^{(j)}, N^{(j)}$ and also the Dirac eigenstates in the $j$th iteration step, 
we define the source terms $T^{(j)}_{\sigma}(r):=T_\sigma[N^{(j)};r], T^{(j)}_{N}(r):=T_N[\sigma^{(j)},N^{(j)};r]$
from \eqref{SET}.
We integrate the Einstein equation \eqref{EinsteinN} and obtain the 
new metric function
\begin{align}
N^{(j+1)}(r)=1-\frac{\alpha^2}{r}\int_0^rT^{(j)}_\sigma(r')dr'\,.
\label{metricfunctionN}
\end{align}
Substituting \eqref{metricfunctionN} into \eqref{EinsteinN}, we get the equation
\begin{align}
&\frac{\sigma'}{\sigma}=-\frac{\alpha^2}{r}\tilde{T}^{(j+1)}_N,~~
\tilde{T}^{(j+1)}_N:=g_1(r)\frac{1}{2\sqrt{N^{(j+1)}(r)}}+g_2(r)\,. 
\label{Einsteins2}
\end{align}
It is easy to integrate \eqref{Einsteins2} and obtain
\begin{align}
&\sigma^{(j+1)}(r)=\exp\biggl[\alpha^2\int_r^\infty \frac{\tilde{T}^{(j+1)}_N(r')}{r'}dr'\biggr]\,,
\label{metricfunctions}
\end{align}
The solution \eqref{metricfunctionN}, \eqref{metricfunctions} again is plugging into the matrix Hamiltonians
\eqref{matnaa}-\eqref{matnbb} and also \eqref{matuaa}-\eqref{matubb}
and solve the eigenproblem.  

The dimensionless version of the ADM mass in the $j$th iteration step is defined by \label{metricfunctionN}
$M'_\textrm{ADM}:=M_\textrm{ADM}/(2 af_\pi N_c)$ 
\begin{align}
&{M'}_\textrm{ADM}^{(j)}:={M'}_\textrm{ADM,val}^{(j)}(r)+{M'}_\textrm{ADM,sea}^{(j)}(r)\,,
\nonumber \\
&{M'}_\textrm{ADM,val}^{(j)}=\lim_{r\to\infty}\frac{1}{2}\int_0^r T^{(j)}_{\sigma,\textrm{val}}(r')dr'\,,
\nonumber \\
&{M'}_\textrm{ADM,sea}^{(j)}=\lim_{r\to\infty}\frac{1}{2}\int_0^r T^{(j)}_{\sigma,\textrm{sea}}(r')dr'\,.
\label{ADMmass}
\end{align}
The ADM mass is obtained once there has been sufficient convergence ($j\to\infty$).

\begin{figure}[t]
	\centering 
	\includegraphics[width=1.0\linewidth]{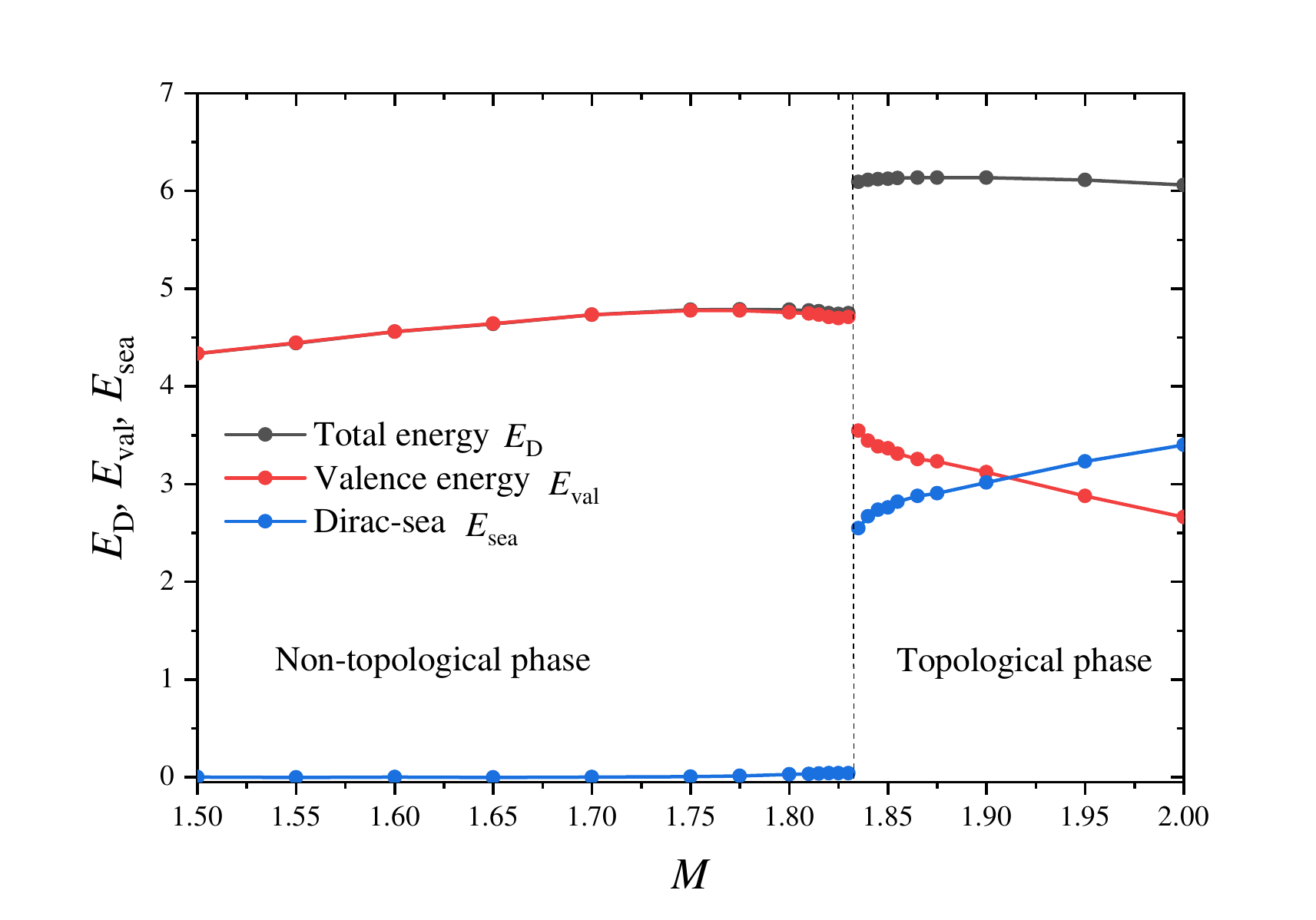}\hspace{-1.0cm}
	\\
	\caption{\label{phasetranstion}The valence, the Dirac-sea and their sum of the energy of the 
	fermions \eqref{fermionenergy} with the low dynamical mass for the 
	coupling constant $\alpha^2=0.4$ is plotted. 
	At the critical $M= M_0\sim 1.835$, a phase transition is observed and above $M_0$, 
	the topological phase emerge, where the chiral symmetry is spontaneously broken.
	At low $M<M_0$, there is no Dirac-sea and only the 
	valence level contributes the energy, which corresponds to the solutions of 
	standard Einstein-Dirac system with the current mass $M$. }
\end{figure}

\begin{figure*}[htbp]
	\centering 
	\includegraphics[width=0.5\linewidth]{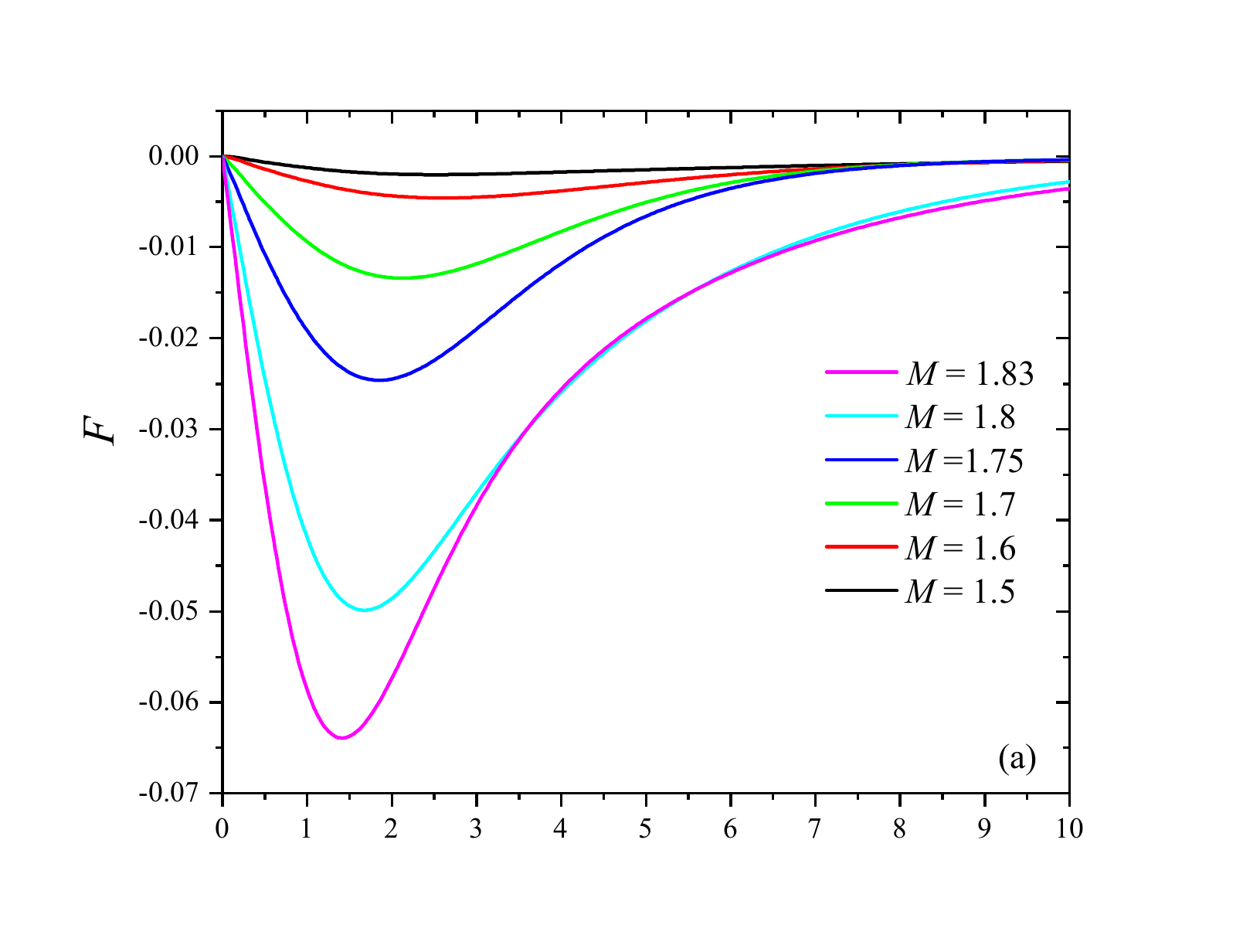}\hspace{-1.0cm}
	\includegraphics[width=0.5\linewidth]{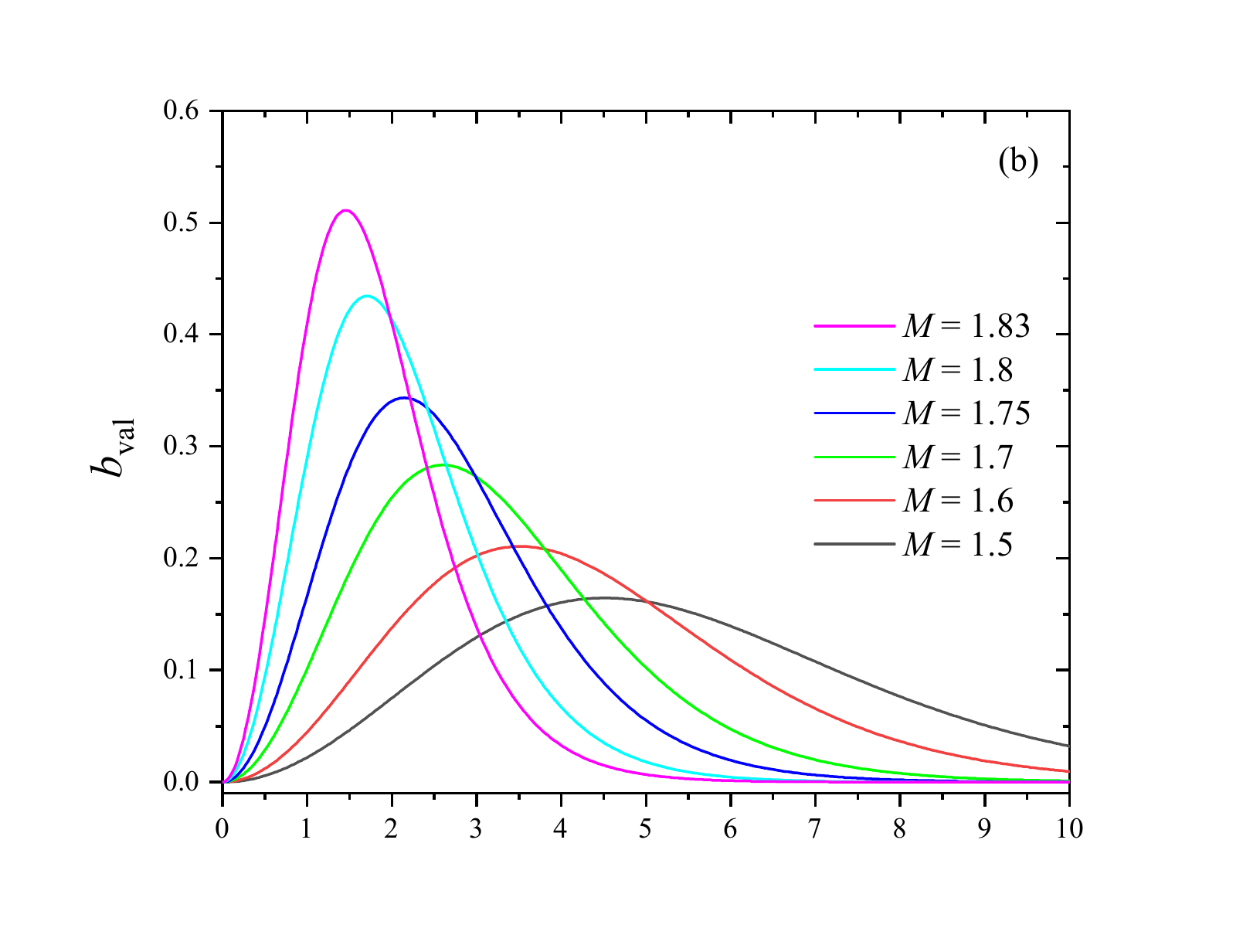}
	\\
	\vspace{-1.0cm}
	
	\includegraphics[width=0.5\linewidth]{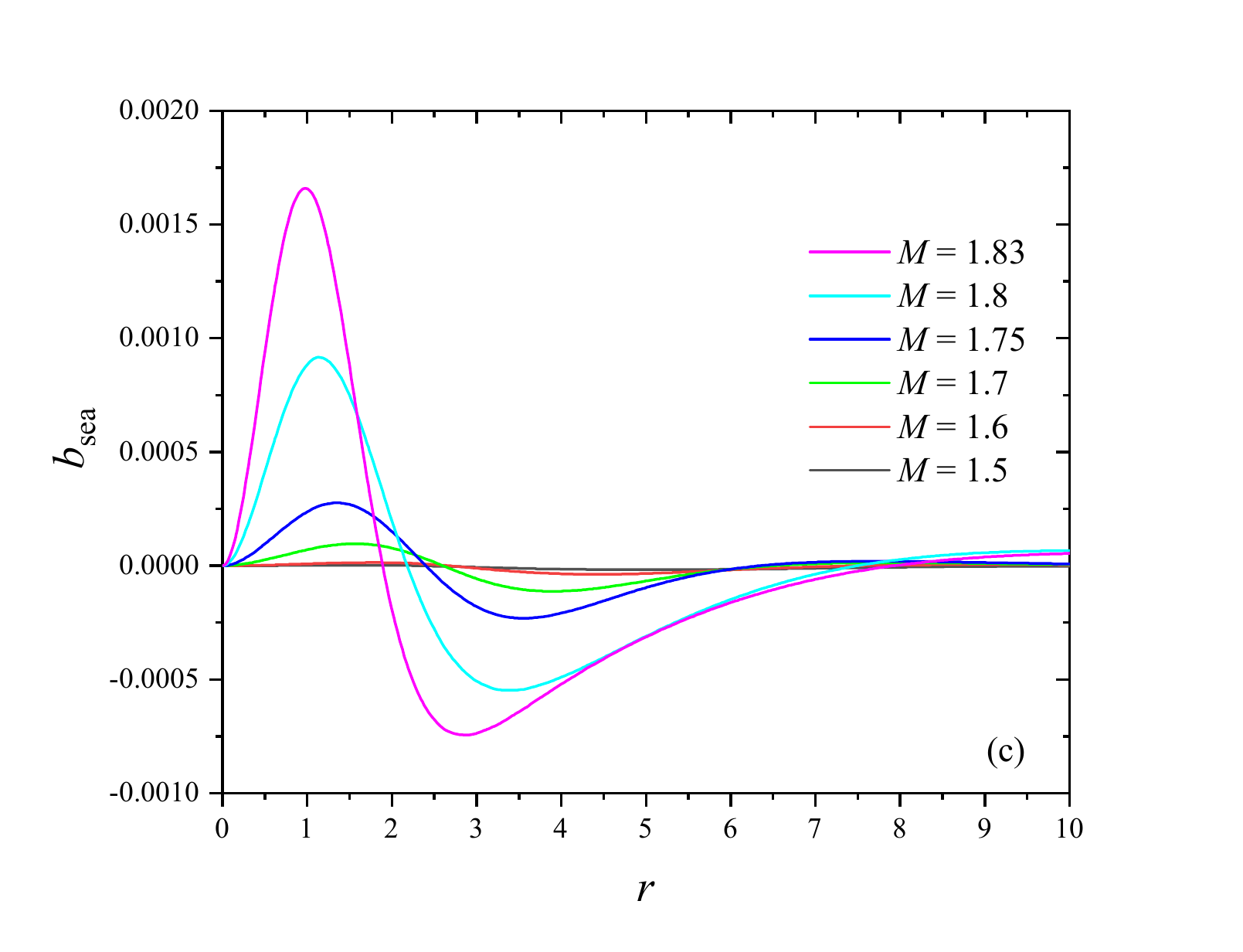}\hspace{-1.0cm}
	\includegraphics[width=0.5\linewidth]{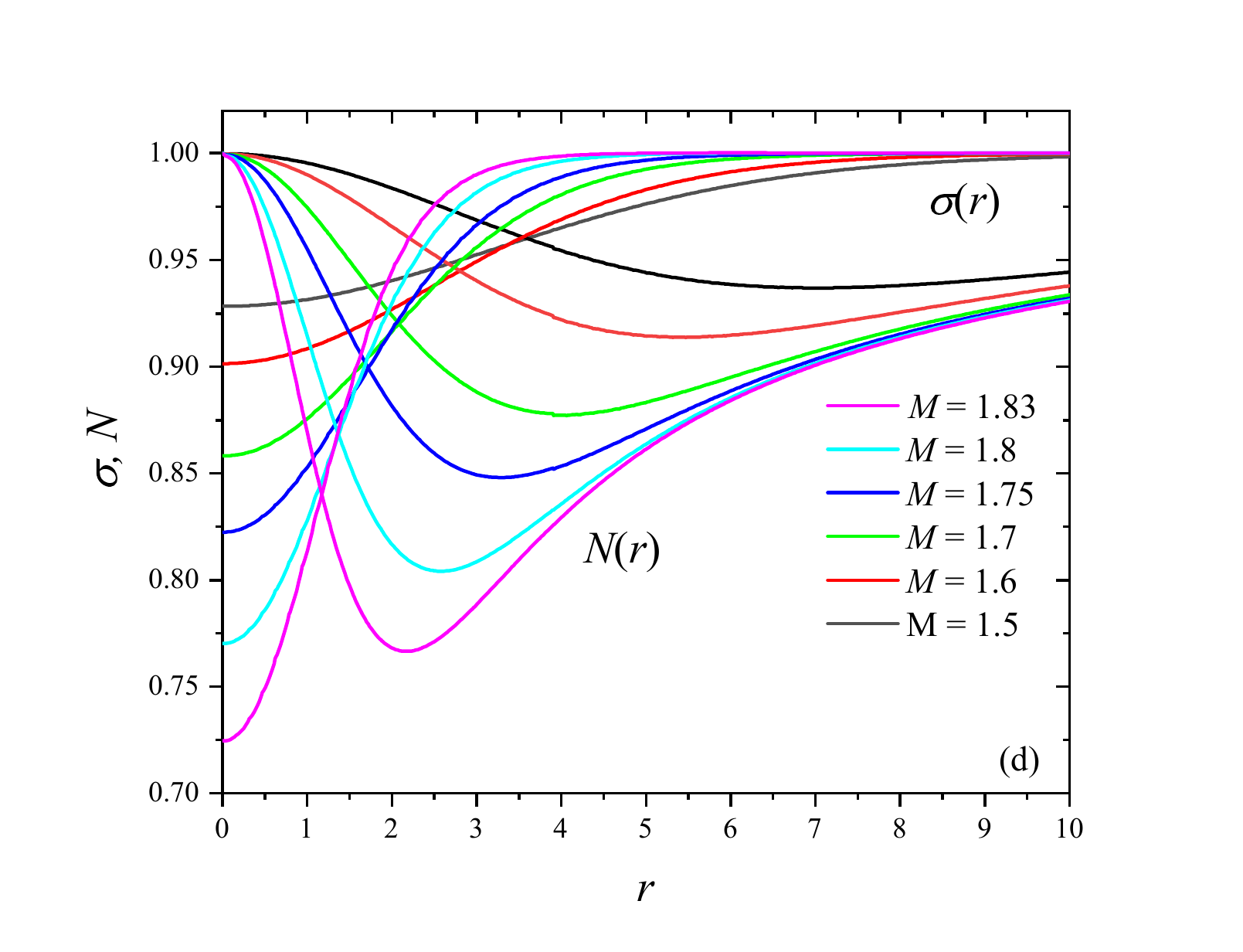}
	
	\caption{\label{Solution_nt}The non-topological soliton solutions of $\alpha^2=0.4$.  
	(a) the profile function $F(r)$, 
	(b) the valence fermion density $b_\textrm{val}(r)$, 
	(c) the Dirac-sea density $b_\textrm{sea}(r)$, and 
	(d) the metric functions $\sigma (r), N(r)$. }
\end{figure*}


\section{\label{sec4}Numerical results}

We first present the result of the flat ($\alpha^2=0$) case as a reference. 
In Fig.~\ref{Solution_flat}, 
we plot the self-consistent profiles $F(r)$ and the valence and the Dirac-sea contribution to the fermion densities $b(r)$ 
defined by~\eqref{fermiondensity},  
with several dynamical mass $M$. The profile function gradually spreads outward as $M$ increases, 
but the fermion density becomes more concentrated. 
In particular, the Dirac-sea contribution $b_\textrm{sea}(r)$ has a negative value, which grows the norm as $M$ increases.
For $M=4.0$, the valence spectrum of the solution dives into the negative energy region, and the Dirac-sea contribution 
has the positive value. 

Our strategy for finding the gravitating solutions, 
is to successively solve different types of equations, such as the Dirac eigensystem, 
the Skyrme equation, and the Einstein equation, and then resolve the results in a self-consistent manner.
In Fig. \ref{Iteration}, we attempt to illustrate the steps involved in our numerical self-consistent analysis. 
A typical iteration process of the total fermion energy $E_D$ \eqref{fermionenergy} 
and 	also the ADM mass $M'_\textrm{ADM}$ \eqref{ADMmass} is shown as a function of the iteration step $\tau$.
There are a few sequences for the convergence. 
The first sequence of the self-consistent analysis \textbf{Step 1} - \textbf{Step 3} 
is $0\leq \tau \leq 89$, 
where the simulation is performed for $\sigma_0(r) =N(r)=1.0$ until the convergence is achieved. 
Then, the second sequence $90\leq \tau \leq 100$ is performed 
for the metric functions at $\tau=90$ evaluated via Eqs.\eqref{metricfunctions}. 
The third sequence is $101  \leq \tau \leq 109$ and the fourth is $108 \leq \tau \leq 109$. 
Though the analysis is somewhat complicated and a little slow for sufficient convergence, it is effective
in identifying the solutions of the present system.

\subsection{The topological phase}

The spectral flow is one of the essential contents of fermions coupled with topological solitons. 
Figure~\ref{Spectral flow} shows the spectral flow of the fermion energies including the valence and the sea 
contributions~\eqref{fermionenergy} as a function of 
the dynamical mass $M$ with the coupling constants $\alpha^2=0,0.1,0.2,0.3,0.4$. Here we employ the 
color number $N_\textrm{c}=3$. It is well-known that for larger $M$, energy of the valence (normalizable) 
fermion tends to dive into negative energy region, which guarantees the coincidence of the valence fermion 
number and the background topological number of the skyrmion. 
The behavior of $E_\textrm{val}$ gets dense and the energy of the Dirac-sea $E_\textrm{sea}$ increases  
with larger coupling constants $\alpha^2$. The total energy $E_\textrm{D}$ always decreases as the dynamical 
mass or the coupling constant grows. 
Consequently, the solutions are more tightly bound for larger $M, \alpha^2$.
In this paper, we primarily examine the region $M:[1.7, 2.5]$ where the ADM mass is positive or 
negative but close to zero. 

Figure \ref{Solution01} plots the numerical results of the profile function $F(r)$, 
the valence and the Dirac-sea part of the fermion density \eqref{fermiondensity} and  
the metric functions $\sigma(r),N(r)$ with several dynamical mass $M$. 
For larger $M$, the fermions and the profiles are confined to the origin and the 
Dirac-sea components gradually enhance. The metrics distort as $M$ increases, 
which is consistent with our notion that gravity is stronger for objects with 
larger masses. 

In Fig.\ref{ADM01}, we plot the ADM mass derived by \eqref{ADMmass} for the solution of Fig.\ref{Solution01}. 
The valence and the sea contributions to the ADM mass $M'_\textrm{ADM,valence},M'_\textrm{ADM,sea}$ 
always appears as opposite sign. The total ADM mass $M'_\textrm{ADM}$ reduces as $M$ increases and 
at some critical point, it tends to be negative value. Therefore, for physically reasonable result, 
suitable value of the dynamical mass is restricted $M\lesssim 2.0$. 
As mentioned previous in sec.II, the dynamical mass is the dimensionless quantity. 
However, assuming the dynamical mass has inverse of length (and also the radius $r$ has the length),
we convert them into the dimensionful quantity using $\hbar c\sim 197.3$ MeV fm, and  
the above condition corresponds to $M\lesssim 395$ MeV. 
The value is shared by majority of the CQSM literature in flat space-time, 
where it is suggested that $370\leq M \leq 420$ MeV in order to get good agreement 
with a number of physical quantities.
Similar behavior of the ADM mass in terms of model for the one fermion and the skyrmions was discussed in 
\cite{Dzhunushaliev:2024kti}. 
Notably, it is physically intriguing that there is an inactive point for the impact of gravity. 
The dynamical mass $M\sim 1.9$, where 
the fermion energy does not change for different $\alpha^2$ and also the ADM mass becomes almost zero. 
As we mentioned, there is a point of zero ADM mass which means quasi-flat $N(r)$ around
$M=1.9$. Later we shall discuss why such a condition could arise.

We compute the source terms of the Einstein equations to observe how the quasi-flat solution appears.
In Fig.\ref{EMtensor01}, where we plot the valence, the sea and the sum of the source term $T_\sigma$~\eqref{EMtensors2}
and $T_N$~\eqref{EMtensorN2}. The sea components always suppress the 
valence because it has the opposite sign of the valence, 
which mechanism is entirely absent when only the valence is taken into account. 
Because both contributions nearly balance one other out, the influence of gravity is minimized at $M\sim 1.9$. 

As a reference, we plot cases for larger $M$ in Fig.\ref{Solution_high}, where the ADM mass is always 
negative.

\subsection{The non-topological phase}

When $M$ is below at some critical point, we get the solution with non-topological chiral field. 
In Fig.\ref{phasetranstion}, we plot the fermion energy of $\alpha^2=0.4$ around the critical value 
$M=M_0\sim 1.835$. 
The energy changes discontinuously, with the Dirac-sea becoming nearly nothing and the valence energy 
suddenly increasing. The total energy transitions to the significantly lower value. 
Above $M_0$, the fermion is localized by the gravity and also the skyrmion which appears on the 
breaking of chiral symmetry.  
The fermion is loosely bound below $M_0$ because only the weak gravity is responsible for it. 
Such sudden changes in the localization and then the energies 
happen because the symmetry breaking is a second-order transition.
Figure \ref{Solution_nt} shows the solutions with $M=1.5, 1.6, 1.7, 1.75, 1.8, 1.83$. 
Apparently, they largely expand as they move away from the critical $M_0$. The Dirac-sea
tends to diminish if the boundary extends. So, it is reasonable 
that if the solutions is in the non-topological phase, 
the Dirac-sea contribution can be negligible. 
The solutions of $M\lesssim M_0$ (i.e. $F(r)\sim 0$) are thus correspond to the ones of the standard 
Einstein-Dirac system with the current mass $M$. 
As $M$ grows, the dynamical fermions appear at $M=M_0$ with the radius shrinkage, 
and then the Dirac-sea contribution will take a dominant role in the solutions.

\begin{figure}[t]
	\centering 
	\includegraphics[width=1.0\linewidth]{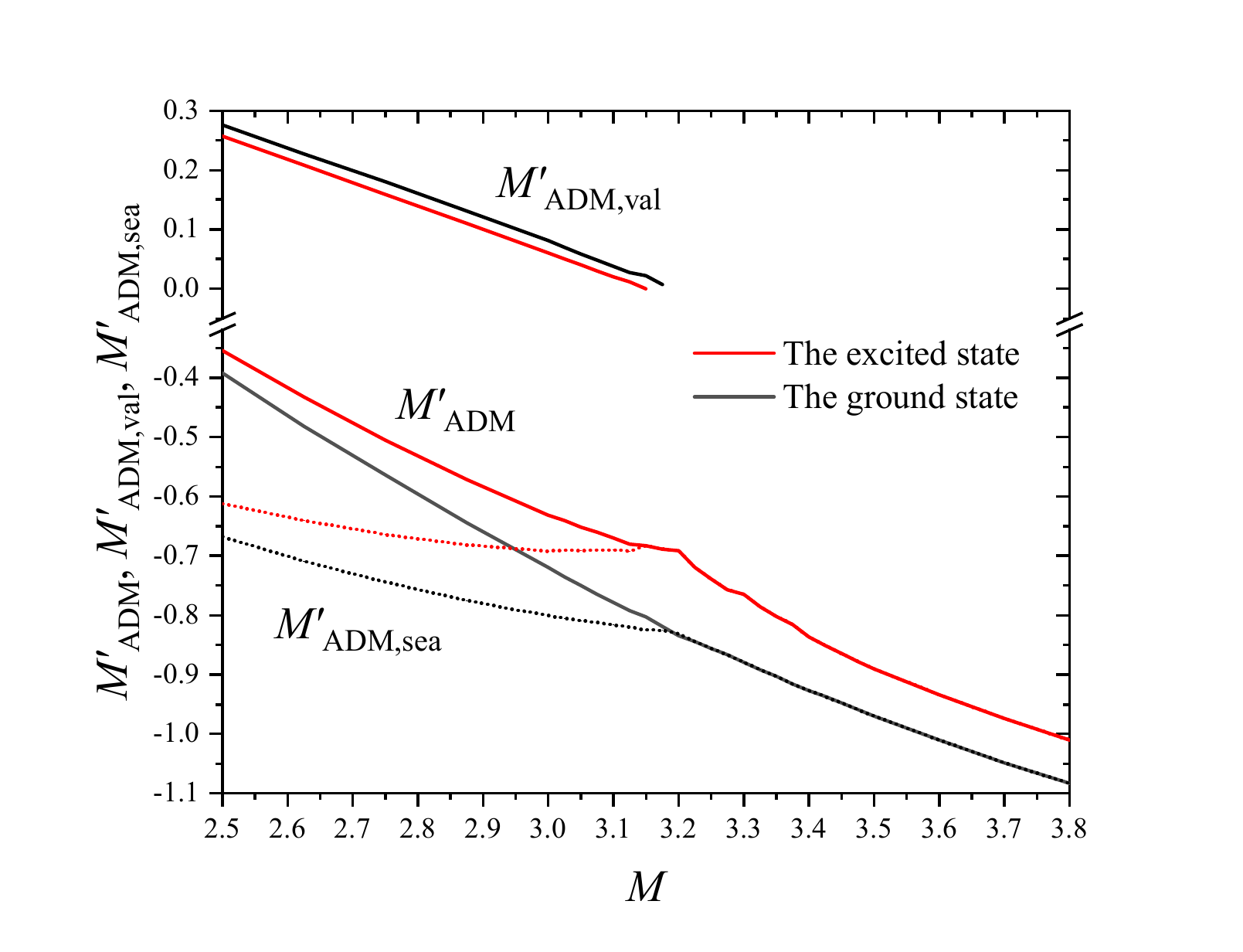}

	\caption{\label{ADMmass_high}
	The ground state and an excited state of the solution. 
	The valence, the Dirac-sea and their sum of the ADM mass 
	with the dynamical mass $M:[2.5,3.8]$ and the gravitational 
	coupling constant $\alpha^2=0.1$ is plotted. Around $M\sim 3.2$, 
	the valence contribution falls into the negative region, and
	only the Dirac-sea component carries the total ADM mass.
	}
\end{figure}

\section{\label{sec5}Summary}

We have developed a new formulation of the Einstein-Dirac system in terms of the chiral quark soliton model.
The model is composed of the valence and the Dirac-sea contributions in a spherically symmetric, 
Schwarzschild-like metrics. The normalizable spinors, the skyrmion profile, and the metric functions 
are all successfully obtained for the several dynamical mass $M$ and the coupling constant $\alpha^2$
 via our self-consistent treatment. Below the critical value of $M$, we also found the solution with 
non-topological chiral field which will connect to Finster's solution in the standard Einstein-Dirac system. 
Our result is some connection with the Einstein-Dirac-Skyrme theory developed in \cite{Dzhunushaliev:2024kti}, 
where the gravitating skyrmion is taken into account as well. They also pointed out there appears the solutions
with negative ADM mass which are also emerged our model caused by excess of the Dirac-sea contribution. 
The most crucial question is how far the ADM negative mass region spans; this depends on the excess of the Dirac-sea, 
that is, the regularization scheme and the cutoff value chosen to regularize the infinite sum of the Dirac eigenstates.

It is widely known that the Einstein-Dirac systems have several different types of excitation branches.
In the present paper,  
we only have discovered the ground state solutions, but  
we also observe that there are some excited states close to the ground state one. 
In Fig.~\ref{ADMmass_high}, we plot the ADM mass of the ground states and the excited states
because the energies are not so visible between the ground/excited states. 
Around $M\sim 3.2$, the difference is obvious and it retains as $M$ increases.
We guess that there are other (unstable) branches from the various valence and Dirac-sea contribution balances 
that our numerical scheme was unable to identify.
Understanding the full structure of the spectrum in the present model requires a more efficient scheme 
to obtain the excited states.

The following open problems should be resolved in future research. 
Our result is the regularization scheme dependent. The proper-time regularization scheme has been mostly used in 
the chiral quark soliton model in flat spacetime. However, in order to verify the validity found in this paper, 
we should employ an alternative scheme, such as the Pauli-Villars method.
The solutions shown in the present paper are totally classical ones. The fermions have the grandspin
which is a mixing of the spin and the flavor. 
In this sense, the solutions possess no proper quantum numbers, and
after an appropriate quantization scheme such as a collective coordinate method is applied, we are able to obtain
the correct observable including the mass, the spin and also the flavors larger than two, carried by fermions. 
At some hot and dense medium, it is expected that a restoration of the chiral symmetry occurs and the property
of the ingredient is drastically modified~\cite{Alkofer:1990if,Christov:1991ry,Berger:1996hc,Nagai:2004jk,Pradhan:2025yml}. 
The extension of our model into the hot dense medium will be efficient for the study of the evolution of 
the early universe or the structure of the neutron stars. 
We shall report on these issues in future studies. 

\begin{center}
{\bf Acknowledgment}
\end{center}

The authors would like to thank Yakov Shnir, Shunsuke Iwasawa and Shintaro Yamamoto for useful advices and comments.  
N.S. was supported in part by JSPS KAKENHI Grant Number JP23K02794.

\appendix

\section{Useful formula for the Hamiltonian matrix elements}

Since the operator $\bm{\sigma}\cdot\hat{\bm{r}}$ has the matrix form
\begin{align}
\bm{\sigma}\cdot\hat{\bm{r}}=
\begin{pmatrix}
\cos\theta & \sin\theta e^{-i\phi} \\
\sin\theta e^{i\phi} & -\cos\theta \\
\end{pmatrix}\,,
\end{align}
we immediately derive the following relation using explicit form of the angular basis
\begin{align}
&\bm{\sigma}\cdot\hat{\bm{r}}|0\rangle =-|2\rangle,~~\bm{\sigma}\cdot\hat{\bm{r}}|2\rangle =-|0\rangle,
\nonumber \\
&\bm{\sigma}\cdot\hat{\bm{r}}|1\rangle =-|3\rangle,~~\bm{\sigma}\cdot\hat{\bm{r}}|3\rangle =-|1\rangle\,.
\label{srsp}
\end{align}
Square of the total angular momentum $\bm{j}=\bm{l}+\frac{\bm{\sigma}}{2}$ leads
$\bm{\sigma}\cdot\bm{l}=\bm{j}^2-\bm{l}^2-\frac{\bm{\sigma}^2}{4}$. 
For the state $|l,j,m\rangle$, we get
\begin{align}
\bm{\sigma}\cdot\bm{l}|l,j,m\rangle=\Bigl(j(j+1)-l(l+1)-\frac{3}{4}\Bigr)|l,j,m\rangle\,.
\end{align}
Therefore, for the angular basis we obtain
\begin{align}
&\bm{\sigma}\cdot\bm{l}|0\rangle =K|0\rangle,~~\bm{\sigma}\cdot\bm{l}|2\rangle =-(K+1)|2\rangle,
\nonumber \\
&\bm{\sigma}\cdot\bm{l}|1\rangle =-(K+1)|1\rangle,~~\bm{\sigma}\cdot\bm{l}|3\rangle =(K-1)|3\rangle\,.
\label{slsp}
\end{align}
The estimation concerning the kinetic term of the Hamiltonian~\eqref{Hamiltonian} 
\begin{align}
&-i(\gamma^0\underline{\gamma}^r\partial_r
+\gamma^0\underline{\gamma}^\theta\partial_\theta
+\gamma^0\underline{\gamma}^\phi\partial_\phi)
\equiv I\otimes(\bm{\sigma}\cdot\hat{\bm{k}})\,,
\nonumber \\
&\hat{k}_x:=-i\biggl(\sqrt{N}\sin\theta\cos\phi\partial_r+\frac{\cos\theta\cos\phi}{r}\partial_\theta-\frac{\sin\phi}{r\sin\theta}\partial_\phi\biggr)\,,
\nonumber \\
&\hat{k}_y:=-i\biggl(\sqrt{N}\sin\theta\sin\phi\partial_r+\frac{\cos\theta\sin\phi}{r}\partial_\theta+\frac{\cos\phi}{r\sin\theta}\partial_\phi\biggr)\,,
\nonumber \\
&\hat{k}_z:=-i\biggl(\sqrt{N}\cos\theta\partial_r-\frac{\sin\theta}{r}\partial_\theta\biggr)
\end{align}
can be analytically performed. 
In fact, it is enough to know how it operates $\bm{\sigma}\cdot\hat{\bm{k}}$ to the upper or the lower spinor 
components. The result is 
\begin{align}
&\bm{\sigma}\cdot\hat{\bm{k}}j_K(kr)|0\rangle
=\frac{(\bm{\sigma}\cdot{\bm{r}})(\bm{\sigma}\cdot{\bm{r}})}{r^2}\bm{\sigma}\cdot\hat{\bm{k}}j_K(kr)|0\rangle
\nonumber \\
&=\frac{\bm{\sigma}\cdot{\bm{r}}}{r^2}\Bigl(\bm{r}\cdot\tilde{\bm{k}}+i\bm{\sigma}\cdot\bm{l}\Bigr)j_K(kr)|0\rangle
\nonumber \\
&=\frac{\bm{\sigma}\cdot{\bm{r}}}{r^2}\Bigl(-ir\sqrt{N}\partial_r+iK\Bigr)j_K(kr)|0\rangle
\nonumber \\
&=\frac{\bm{\sigma}\cdot{\bm{r}}}{r^2}ikr\biggl\{(1-\sqrt{N})\frac{K}{kr}j_K(kr)+\sqrt{N}j_{k+1}(kr)\biggr\}|0\rangle
\nonumber \\
&=-ik\biggl\{(1-\sqrt{N})\frac{K}{kr}j_K(kr)+\sqrt{N}j_{K+1}(kr)\biggr\}|2\rangle
\end{align}
where we have used \eqref{srsp}, \eqref{slsp} and also the well-known recursive formula for the spherical Bessel 
function~\cite{abramowitz1965handbook}
\begin{align}
&\frac{d}{d\rho}j_\ell(\rho)=j_{\ell-1}(\rho)-\frac{\ell+1}{\rho}j_\ell(\rho)
=-j_{\ell+1}(\rho)+\frac{\ell}{\rho}j_\ell(\rho), 
\nonumber \\
&j_{\ell-1}(\rho)+j_{\ell+1}(\rho)=\frac{2\ell+1}{\rho}j_\ell(\rho)\,.
\nonumber 
\end{align}
Similarly, we can get
\begin{align}
&\bm{\sigma}\cdot\hat{\bm{k}}j_K(kr)|1\rangle
\nonumber \\
&=ik\biggl\{(1-\sqrt{N})\frac{K+1}{kr}j_K(kr)+\sqrt{N}j_{K-1}(kr)\biggr\}|3\rangle
\nonumber \\
&\bm{\sigma}\cdot\hat{\bm{k}}j_K(kr)|2\rangle
\nonumber \\
&=ik\biggl\{(1-\sqrt{N})\frac{K+2}{kr}j_{K+1}(kr)+\sqrt{N}j_{K}(kr)\biggr\}|0\rangle
\nonumber \\
&\bm{\sigma}\cdot\hat{\bm{k}}j_K(kr)|3\rangle
\nonumber \\
&=-ik\biggl\{(1-\sqrt{N})\frac{K-1}{kr}j_K(kr)+\sqrt{N}j_{K}(kr)\biggr\}|1\rangle\,.
\end{align}
We successfully deduce the matrix form of the Hamiltonian~\eqref{matnaa}-\eqref{matubb} 
using all these formalisms.

	\bibliography{CQSM}

\end{document}